\documentclass[review]{elsarticle}
\usepackage{lineno,hyperref}
\modulolinenumbers[5]
\usepackage{comment}
\usepackage{amssymb}
\usepackage{amsmath}
\usepackage{footmisc}
\usepackage{xcolor}
\usepackage[utf8]{inputenc}
\usepackage{ulem}
\usepackage{bbold}
\DeclareUnicodeCharacter{202F}{\,}
\newcommand{\jk}[1]{\textcolor{black}{#1}}
\newcommand{\sh}[1]{\textcolor{black}{#1}}

\newcommand{\rp}[1]{\textcolor{black}{#1}}
\begin{document}
\begin{frontmatter}

\title{Majorana Fermions in Magnetic Chains}
%\tnotetext[mytitlenote]{Fully documented templates are available in the elsarticle package on %\href{http://www.ctan.org/tex-archive/macros/latex/contrib/elsarticle}{CTAN}.}

\author[Basel]{R{\'e}my Pawlak\corref{cor1}}
\cortext[cor1]{Corresponding author}
\author[Basel]{Silas Hoffman}
\author[Basel]{Jelena Klinovaja}
\author[Basel]{Daniel Loss}
\author[Basel]{Ernst Meyer}

\address[Basel]{University of Basel, Department of Physics, Klingelbergstr. 82 CH-4056 Basel, Switzerland}

\begin{abstract}
Majorana fermions have recently garnered a great attention outside the field of particle physics, in condensed matter physics. In contrast to their particle physics counterparts, Majorana fermions are zero energy, chargeless, spinless, composite quasiparticles, residing at the boundaries of so-called topological superconductors. Furthermore, in opposition to any particles in the standard model, Majorana fermions in solid-state systems obey non-Abelian exchange statistics that make them attractive candidates for decoherence-free implementations of quantum computers. In this review, we report on the recent advances to realize synthetic topological superconductors supporting Majorana fermions with an emphasis on chains of magnetic impurities on the surface of superconductors. After outlining the  theoretical underpinning responsible for the formation of Majorana fermions,
we report on the subsequent experimental efforts to build topological superconductors and the resulting evidence in favor of Majorana fermions, focusing on scanning tunneling microscopy and the hunt for zero-bias peaks in the measured current. We conclude by summarizing the open questions in the field and propose possible experimental measurements to answer them.
 \\
 
\end{abstract}


\end{frontmatter}

%\linenumbers

\section{The search for the Majorana \jk{fermions}}%fundamental particles}

In an attempt to find a Lorentz covariant quantum equation of motion for the electron, P. A. M. Dirac discovered an equation which today bears his name: the Dirac equation \cite{dirac28}. Upon solving his eponymous equation, Dirac found that electrons are \textit{complex}-valued fields and, inherent to his equation, was the prediction that electrons have an oppositely charged antiparticle partners dubbed positrons. Several years later, Ettore Majorana found a solution to the Dirac equation by imposing the condition of \textit{reality} on his solution \cite{majorana37}. A direct consequence of the reality of this Majorana solution is that the particle does not couple to the electromagnetic field and that it is its own antiparticle. Due to the former property, so-called Majorana fermions were proposed as potential candidates for neutrinos and for cold dark matter. However, using all the power of all the particle accelerators in the world, the existence of Majorana fermion as fundamental particle in the Standard Model is yet dubious. 

Nonetheless the mystique of the Majorana fermion has transcended the realm of particle physics. Although condensed matter physics typical works with energy scales much smaller than particle physics, of the order $1$~eV and smaller, considerable effort has been made to find Majorana fermions composed of electrons and bound by the material properties of specific solid state systems. Typically, the underlying material is a semiconductor or metal with a well-defined Fermi surface. One can excite the system by either adding an electron above the Fermi surface or removing an electron below it, the latter is called a hole. Analogous to the particle-antiparticle relation that the electrons and positrons have in particle physics, electrons and holes hold the same effective relation in condensed matter physics. Consequentially, an equal superposition of an electron and its hole partner with the same spin is by definition a fermion which is its own antiparticle, a Majorana fermion. As such, the search for a solid state system capable of supporting such a stable state is at the heart of the search for Majorana fermions in the condensed matter systems. 
%\footnote{\sh{The interested reader may find an excellent introduction to Majorana fermions in both subfields of physics in Ref.~\cite{Chamon2010,Beenakker2014,Elliott2015}.}}.

A natural starting point is within a conventional, $s$-wave superconductor, described  by Bardeen-Cooper-Schrieffer (BCS) theory \cite{bardeenPR57}.  At sufficiently low temperature in metals and in the presence of lattice vibrations giving rise to phonons that cause attraction between electrons,  electrons of opposite spin are energetically favored to form  bosonic bound states known as Cooper pairs, which Bose-Einstein condense to a superconducting ground state. The effective interaction between electrons can be accommodated using a mean field theory approach. Although such an approximation reduces a rather complex interacting theory to a quadratic Hamiltonian, the conservation of particle number is sacrificed and only the total fermion parity is conserved, i.e. particle number modulo two. As a direct consequence, the excitations, known as Bogoliubov quasiparticles, are superpositions of electrons and holes \jk{with the opposite spin}. One can show that, generally, a quasiparticle at zero energy has equal parts of electron and hole content and is therefore chargeless but still has a finite spin and therefore cannot be its own antiparticle \footnote{\jk{Foreshadowing our discussion of states localized to magnetic impurities in $s$-wave superconductors in Sec.~\ref{shiba}, one can show that finely turning the exchange interaction of the impurity with the quasiparticles in the superconductor can induce a zero-energy and chargeless but spinful localized fermionic state which is not a Majorana fermion \cite{yuAPS65,shibaPTP68,rusinovJETP69b}.}}. In order to obtain a Majorana fermion quasiparticle with zero spin, the spin of the composite electron and hole \jk{must negate each other} %must be the same 
which requires a more exotic pairing in which electrons with the same spin form a Cooper pair and condense giving rise to so-called $p$-wave superconductivity. As such a pairing is exceedingly rare in Nature, hybrid superconducting-semiconducting systems in which spin degeneracy is lifted by an effective inhomogenous magnetic fields, are used to mimic such a spinful pairing in superconductors.  

Before proceeding to outline the review, we give a basic introduction to the properties of Majorana fermions in solid-state systems, as compared with particle physics, which simultaneously presents us with the opportunity to familiarize the reader with vocabulary common in the current condensed-matter literature. While such Majorana fermion quasiparticles can exist in higher dimensions, here we restrict our consideration to the case where they are localized to the boundaries of one-dimensional (1D) or quasi-one-dimensional topological superconductors. \jk{That is, throughout this review, we specifically consider localized nonpropogating Majorana fermions and refer to other literature focusing on delocalized propagating Majorana fermions which can, for instance, be found on the boundary of two-dimensional topological superconductors \cite{beenakkerARCMP13}.}  Topological superconductors are so named because they are superconductors with a topologically nontrivially mapping from momentum space to eigenstates of the Hamiltonian. This property guarantees the existence of zero-energy Majorana fermion states at the end of the topological superconductor; as such they are typically referred to as Majorana zero modes (MZMs), which we practice for the remainder of the manuscript. Furthermore, since they are protected by the topological nature of the system, they are robust to local electric or magnetic noise, which is also a consequence of the fact that they are both chargeless and spinless. Occurring at the boundaries of 1D topological superconductors, the MZMs always come in pairs. \jk{Lastly, due to their spinor representation, their 
braiding or exchange statistics is non-Abelian in the following sense.}
\jk{ The  many-body ground state of several MZMs is degenerate and the many-body spinor wavefunctions can be transformed into each other by exchanging the positions of the MZMs. Furthermore, the order in which the MZMs are exchanged will result in inequivalent ground states, similar to the action of multiplication of matrices which, in general, is non-Abelian if the matrices do not commute with each other. This dependence of the degenerate ground state on the order in which MZMs are exchanged  is a manifestation of non-Abelian statistics.} As this property allows for \jk{the fault-tolerant} operation of some gates used in quantum computation and their robustness against electric and magnetic disorder, MZMs have been proposed as an attractive building block for topological quantum bits (qubits).

\sh{Before moving further, one should emphasize again that MZMs in solid-state setups are only distant cousins of Majorna fermions in particle physics. For example, non-Abelian statistics, the property that attracts so much attention to MZMs, is a feature of low-dimensinal systems and is absent for three-dimensional Majorana fermions of particle physics. At the same time, the question of zero or non-zero mass of Majorana fermions, does not arise in solid-state systems, as MZMs are localized bound states for which one cannot define mass. For detailed comparisons between Majorana fermions in solid-state and particle physics, we refer to reviews dedicated to this topic, see  Ref.~\cite{Elliott2015} as well as  Refs. \cite{Chamon2010,Beenakker2014}}.

Here, we focus on one promising platform for topological superconductivity that appears to have all the ingredients necessary to favor the formation of MZMs: at the ends of a chain of classical spins on the surface of a superconductor. After an obligatory introduction to a minimal  model capable of realizing MZMs, we detail the theoretical underpinnings that support MZMs in spin chains on the surface of a superconductor.  Next, we tour the experiments that have attempted to measure the signature of the MZMs and  subsequently compare and contrast their interpretations; this naturally leads to a discussion of the open questions in the field for which we propose experimental setups to provide possible answers. 

\section{A toy model}
\label{Kitaevsection}
%%%%%%%%%%%%%%%%%%%%%%%%%%%%%%%%%%%%%%%%%%
\begin{figure*}[t!]
	\centering
	\includegraphics[width =0.8\textwidth]{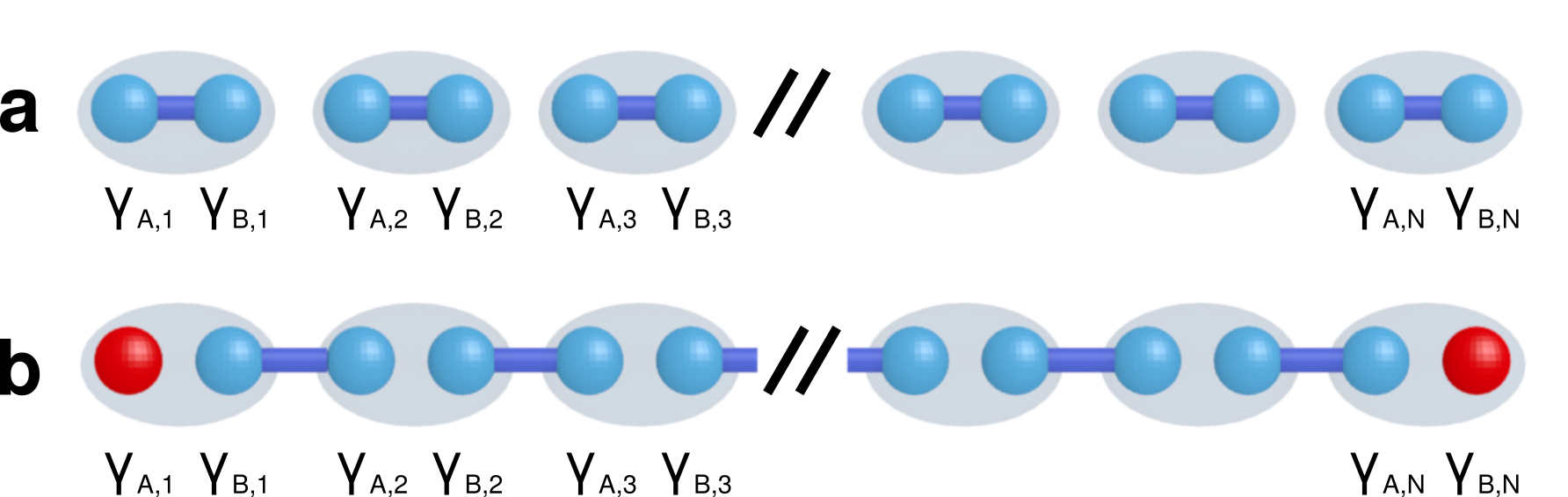}
	\caption{{\bf Kitaev toy model of 1D-spinless $p$-wave superconductors.} {\bf a,} Trivial phase. {\bf b,} Topological phase leading to unpaired Majorana operators at the chain ends, $\gamma_{A,1}$ and $\gamma_{B,N}$ colored in red.}
	\label{FigKitaev}
\end{figure*}
%%%%%%%%%%%%%%%%%%%%%%%%%%%%%%%%%%%%%%%%%%%%%%%%%%%%%%%%%%%%%%%%%%%%%%%%%%%%%%%%%%%%%%%

A simple, illustrative, toy model that realizes MZMs is the Kitaev model \cite{Kitaev2001,Alicea2012},
\begin{eqnarray}
H =  \sum_{j=1}^{N-1} \left[-t c^{\dagger}_{j+1}c_{j} + \Delta c_{j}c_{j+1} + \textrm{H.c}\right] - \mu \sum_{j=1}^{N} c^{\dagger}_{j} c_{j}\,.
\label{Kitaev}
\end{eqnarray}
Here, $c_j^\dagger$ ($c_j$) create (destroy) spinless fermions at site $j$ of an $N$-site chain. The hopping between adjacent sites is of strength $t$, the amplitude of the nearest-neighbor superconducting pairing is $\Delta$, and  $\mu$ is the chemical potential. \jk{We note that as there is no spin degree of freedom, $s$-wave superconductivity [3], which is the singlet pairing of an electron and hole, is absent and the superconductivity in this model, which pairs fermions at adjacent sites rather than on the same site, enforces an unconventional $p$-wave superconductivity.}

A convenience of this model is that one can explicitly obtain zero-energy solutions corresponding to MZMs. Formally, each fermionic operator can be written as the sum of two Majorana operators
\begin{equation}
c_j=\frac{\gamma_{B,j}+i\gamma_{A,j}}{2}\,,\ \ \ c_j^\dagger=\frac{\gamma_{B,j}-i\gamma_{A,j}}{2}.
\end{equation}
Here, $\gamma_{\alpha,j}$ are Majorana operators describing particles that are their own antiparticles, $\gamma_{\alpha, j} = \gamma^{\dagger}_{\alpha, j}$. In addition, these operators anticommute with each other:  $\{\gamma_{\alpha, i},\gamma_{\beta, j}\}$=$2\delta_{\alpha\beta}\delta_{i j}$, for $\alpha,\beta=A,B$. 

First, if $\Delta=t=0$, the chain consists of uncoupled sites, so the system is obviously in the trivial phase, see Fig.~\ref{FigKitaev}(a).
If $\mu=0$ and $\Delta=t$, Eq.~(\ref{Kitaev}) dramatically simplifies to a sum of products of Majorana operators
\begin{equation}
H = - i \frac{t}{2}\sum_{j=1}^{N-1} \gamma_{B,j}\gamma_{A,j+1}.
\label{Kitaev2}
\end{equation}
Such pairing of Majorana operators from neighboring sites is schematically depicted in Fig.~\ref{FigKitaev}(b).
Next, it is natural to define a second set of complex fermion operators composed of Majorana operators from adjacent sites
\begin{equation}
d_j = \frac{ \gamma_{A,j} + i \gamma_{B,j+1}}{2}\,,
\label{Kitaev3}
\end{equation}
wherein the Hamiltonian becomes
\begin{equation}
H = t \sum_{j=1}^{N-1} (d^{\dagger}_{j}d_j - \frac{1}{2})\,.
\label{Kitaev4}
\end{equation}
This Hamiltonian describes a flat band in which the cost to occupy a fermionic state on any site is given by $t$. Two Majorana operators are noticeably absent from Eq.~(\ref{Kitaev2}), $\gamma_{A,1}$ and $\gamma_{B,N}$, as depicted in Fig.~\ref{FigKitaev}(b). As such, these two unpaired Majorana operators, obviously, commute with the Hamiltonian and, therefore, the corresponding excitation resides at zero energy, providing our first explicit solution for MZMs. These two unpaired MZMs located at the ends of the chain can be formally combined into a nonlocal complex fermion,
\begin{eqnarray}
d_0 = \frac{1}{2} (\gamma_{A1}+i \gamma_{BN})\,,
\label{Kitaev5}
\end{eqnarray}
with the usual anticommutation properties. As $[H,d_0]=0$, $d_0$ corresponds to a zero energy state.
As a result, the many-body ground state of the Kitaev chain [see Eq.~(\ref{Kitaev4})] is doubly degenerate corresponding to the occupancy or vacancy of the complex fermionic state filled by $d_0^\dagger$.

One may show that the system is in a topological phase \cite{Kitaev2001} when the chemical potential $\mu$ lies within the band and therefore supports MZMs at the ends. When $|\mu|>2t$, the chain is either completely empty or completely filled for any value of the superconducting pairing; the trivially insulating phase is smoothly connected to the insulating phase with superconducting pairing. Therefore, the system is topologically trivial, \textit{i.e.} absent of MZMs, for any value of $\Delta$ when $|\mu|>2t$. Away from the specific point in parameter space discussed above, $\mu=0$ and $t=\Delta$, the MZMs are no longer localized to a single site but are exponentially localized to the ends of the chain with decay length given by the superconducting coherence length, $\xi \propto 1/\Delta$, \sh{in the regime of interest $\Delta \ll t$}. When the size of the chain is comparable to $\xi$, the MZMs hybridize and split away from zero, lifting the double degeneracy of the ground state \cite{prada1,rainis}.

%One may show  that the system is in a topological phase \cite{Kitaev2001}, \textit{i.e.} supports MZMs, when the chemical potential $\mu$ lies within the band, $|\mu|<2t$, and in the trivial phase, \textit{i.e.} no MZMs at the ends, otherwise (note that $|\mu|>2t$ is actually trivial since it means that the chain is either completely empty or filled; in this case, even superconductivity is not possible). Away from the specific point in parameter space discussed above, $\mu=0$ and $t=\Delta$, the MZMs are no longer localized to single site but are exponentially localized to the ends of the chain with decay length given by the superconducting coherence length, $\xi \propto 1/\Delta$. When the size of the chain is comparable to $\xi$, the MZMs hybridize and split away from zero, lifting the double degeneracy of Eq.~(\ref{Kitaev4}).

\section{Chains of magnetic impurities: Theory}

\jk{Chains of magnetic impurities on the surface of an $s$-wave superconductor are one physical system that appears to have the necessary ingredients to realize MZMs at its ends. Despite the fact that several experiments have discovered zero energy end states in precisely these systems, the mechanism by which the states are conceived is yet in dispute. Firstly, it is unknown if the 1D system furnishing the MZMs at its ends is induced within the bulk superconductor \cite{nadj-pergePRB13,pientkaPRB13,liPRB14,pientkaPRB14,glazovPRB14,heimesPRB14,hoffmanPRB16,poyhonenPRB16,andolinaPRB17,theilerCM18} or realized within the atomic chain  itself  \cite{klinovajaPRL13,vazifehPRL13,brauneckerPRL13}. When the magnetic exchange between an atom and the quasiparticles in the bulk superconductor is sufficiently strong, a localized Yu-Shiba-Rusinov state \cite{yuAPS65,shibaPTP68,rusinovJETP69b} with energy within the superconducting gap is formed. A chain of such magnetic atoms creates many localized states that can hybridize and form a band, which can itself support MZMs at the ends of the chain within the superconductor \cite{nadj-pergePRB13,pientkaPRB13,liPRB14,pientkaPRB14,glazovPRB14,heimesPRB14,hoffmanPRB16,poyhonenPRB16,andolinaPRB17,theilerCM18}.  Alternatively, if the atoms are sufficiently close together, their orbitals can overlap and the system appears as a 1D quantum wire with proximity-induced superconductivity \cite{klinovajaPRL13,vazifehPRL13,brauneckerPRL13}. The mechanism by which the MZM is generated will affect the qualitative properties of the wavefunction such as their localization lengths as well as the period of oscillation.}

\jk{The other unknown is how $p$-wave superconductivity is generated from conventional $s$-wave superconductivity. It is well established that a Zeeman splitting, induced by the exchange interaction with impurities in this case, is a necessary ingredient but the magnetic order of the chain is unknown. Interaction between the magnetic impurities mediated by the quasiparticles of the superconductor induces a helical order with a pitch of $2 k_F$ \cite{klinovajaPRL13,vazifehPRL13,brauneckerPRL13,hsu}. Such a helical order is sufficient to drive the system into the topological phase. However, as the atoms are placed on the surface of a superconductor, a magnetic anisotropy can arise which overwhelms the helical order and forces the polarization perpendicular to the surface of the superconductor. This ferromagnetic order can induce MZMs only if there is a Rashba spin-orbit interaction in the material and the chemical potential lies close to the corresponding spin-orbit crossing point at zero momentum. In contrast to that, the helical ordering results in the self-tuning topological phase independent of the presence of the SOI and occurs at any position of the chemical potential in the chain \cite{klinovajaPRL13}.}

\jk{In this section, we outline the calculations that support MZMs realized by the hybridization of localized states in the superconductor [Sec.~\ref{shiba}] and effective 1D quantum wires. In both cases, we focus on a helical order of the magnetic impurities noting that ferromagnetic order with spin-orbit interaction is equivalent to the former up to a local gauge transformation. We leave a discussion of the factors contributing to the magnetic order and the effects on the formation of MZMs for Sec.~\ref{order}}

\subsection{Hybridization of bound states within a superconductor}
\label{shiba}
Before we discuss a model for a \textit{chain} of magnetic impurities, we recall the properties of a \textit{single} magnetic impurity within a conventional superconductor. It is known that a single magnetic impurity in contact with a superconductor induces a spin-polarized bound state located under the impurity known as a Yu-Shiba-Rusinov state  \cite{yuAPS65,shibaPTP68,rusinovJETP69b}  or, more concisely, a Shiba state. The Bogliobov-de-Gennes Hamiltonian describes a bulk superconductor:
\begin{equation}
H_0=\sum_{\sigma=\uparrow\downarrow}\int d\textbf{r}\Psi_\sigma^\dagger(\textbf{r})\left[-\frac{\hbar^2\nabla^2}{2m}-\mu\right]\Psi_\sigma+ \left[\frac{i\Delta}{2}\Psi_\sigma(\textbf{r})\sigma^y_{\sigma\sigma'}\Psi_{\sigma'}(\textbf{r})+\textrm{H.c.}\right]\,.
\end{equation}
The first term describes a normal metal held at a chemical potential $\mu$, while the second term describes the superconducting pairing between electrons into Cooper pairs with a superconducting order parameter $\Delta$. Here, $\Psi_\sigma^\dagger(\textbf{r})$ [$\Psi_\sigma(\textbf{r})$] creates (annihilates) an electron with spin $\sigma$ and an effective mass $m$ at a space point $\textbf{r}$. A single impurity at  $\textbf{r}=0$ of the size much smaller than the Fermi wavelength can be simply modeled as
\begin{equation}
H_I=-\Psi_\sigma^\dagger(\textbf{r})J\textbf{S}\cdot\boldsymbol\sigma_{\sigma\sigma'}\Psi_{\sigma'}(\textbf{r})\delta(\textbf{r})\,,
\end{equation}
where $J$ is the  strength of the magnetic exchange interaction between the magnetic impurity  and the quasiparticles. In what follows, $\textbf S$=$S$($\cos\phi\sin\theta$, $\sin\phi\sin\theta$, $\cos\theta$) is the magnetic moment of the impurity which is treated as classical vector of length $S$ characterized by polar angles $\phi$ and $\theta$. 

%\jk{Majorana fermions to MZMs}

It is straightforward to find solutions to the full Hamiltonian, $H_S=H_0+H_I$, with energies within the superconducting gap \cite{yuAPS65,shibaPTP68,rusinovJETP69b}:
\begin{equation}
\frac{E_\pm}{\Delta}=\pm\frac{1-\alpha^2}{1+\alpha^2}\,,
\label{E_shiba}
\end{equation}
where $\alpha = \pi \nu_F J S$, $\nu_F$ is the density of states at the Fermi energy $E_F$ (chemical potential $\mu$ at $T=0$).
%and $S=|\textbf S|$. 
The energies of the Shiba states given by Eq.~\ref{E_shiba}, owing to the inherent particle-hole symmetry of the Hamiltonian, are symmetric around the chemical potential. As $\alpha\rightarrow 0$ or $\alpha\rightarrow \infty$, the Shiba state energy approaches the edge of the superconducting gap.
It is exactly in the middle of the gap at $E=0$ if $\alpha=1$. At the point of the impurity, $\textbf{r}=0$, the corresponding modes are given by spinors
\begin{equation}
\Psi_+(\textbf{0})=\left(\begin{array}{c}|\uparrow\rangle \\|\uparrow\rangle
\end{array}\right)\cdot\Psi(\textbf 0)\,,\,\,
\Psi_-(\textbf{0})=\left(\begin{array}{c}  |\downarrow\rangle\\- |\downarrow\rangle
\end{array}\right)\cdot\Psi(\textbf 0)\,,\,\,
\label{basis1}
\end{equation}
where
\begin{align}
 |\uparrow\rangle = \left(\begin{array}{c}\cos\theta/2\\e^{i\phi}\sin\theta/2
\end{array}\right)\,,\,\,
 |\downarrow\rangle = \left(\begin{array}{c}e^{-i\phi}\sin\theta/2\\-\cos\theta/2
\end{array}\right)\, .
\label{spin1}
\end{align}
The spinors are written in the basis $\Psi(\textbf{r})=[\Psi_\uparrow(\textbf{r}),\Psi^\dagger_\downarrow(\textbf{r}),\Psi_
\downarrow(\textbf{r}),-\Psi^\dagger_\uparrow(\textbf{r})]$. In the absence of the spin-orbit interaction, the Shiba state is spin-polarized parallel or antiparallel to the magnetic moment of the impurity. The corresponding wavefunction is a superposition of electrons and holes. It is these two properties that make the Shiba state a good starting point to hunt for MZMs. %\jk{What do you mean here? Spin is always a good quantum number}

%\jk{Use everywhere spin-orbit interaction}

It is straightforward to generalize $H_S$ to a chain of $N$ magnetic impurities located $\textbf{r}_i$ as $H_C=H_0+\sum^N_{i=1}H_i$, where 
\begin{equation}
 H_i=- \Psi_\sigma^\dagger(\textbf{r})J_i\textbf{S}_i\cdot\boldsymbol\sigma_{\sigma\sigma'}\Psi_{\sigma'}(\textbf{r})\delta(\textbf{r}_i-\textbf{r})\,.
\end{equation}
\jk{In the following we consider identical magnetic impurities,  $J_i=J$ and $|\textbf{S}_i|=S$, polarized along different directions, $\textbf{S}_i=(
\cos\phi_i\sin\theta_i,\sin\phi_i\sin\theta_i,\cos\theta_i)$.  In the absence of the spin-orbit interaction, it is necessary that the chain forms a helical structure in order to support MZMs. As such, we consider a planar helix with pitch $k_h$ so that $\theta_i=\pi/2$ and $\phi_i=2k_h r_i$ parameterize the magnetic moment at $\textbf r_i$.} Although it is difficult to find solutions of in-gap states in general, the problem can be considerably simplified if one focuses on the uncoupled Shiba states are near the chemical potential (at zero energy), $\alpha\approx1$, and, if one focuses on the impurities that are sufficiently separated, the so-called dilute limit. This facilitates a projection onto a basis of the uncoupled Shiba states at the sites of the impurities \cite{pientkaPRB14}
\begin{equation}
\Psi_{+i}=\left(\begin{array}{c}|\uparrow, i\rangle \\|\uparrow, i\rangle
\end{array}\right)\cdot\Psi(\textbf{r}_i)\,,\,\,
\Psi_{-i}=\left(\begin{array}{c}  |\downarrow, i\rangle\\- |\downarrow, i\rangle
\end{array}\right)\cdot\Psi(\textbf{r}_i)\,,\,\,
\label{basis2}
\end{equation}
where
\begin{align}
 |\uparrow,i\rangle = \left(\begin{array}{c}\cos\theta_i/2\\e^{i\phi_i}\sin\theta_i/2
\end{array}\right)\,,\,\,
 |\downarrow,i\rangle = \left(\begin{array}{c}e^{-i\phi_i}\sin\theta_i/2\\-\cos\theta_i/2
\end{array}\right)\,.
\label{spin2}
\end{align}
Noticing that $\Psi_{+i}=\Psi_{-i}^\dagger$, the resulting effective Hamiltonian is given by  \cite{pientkaPRB14}
\begin{equation}
H_{\textrm{eff}}=\sum_{i=1}^N E_+ \Psi_{+i}^\dagger\Psi_{+i} +\sum_{i\neq j}\left
(h^{\textrm{eff}}_{ij}\Psi_{+i}^\dagger\Psi_{+j}+\Delta^{\textrm{eff}}_{ij}\Psi_{+i}\Psi_{+j} +\textrm{H.c.}\right)\,,
\label{Heff}
\end{equation}
where we introduce notations
\begin{align}
h^{\textrm{eff}}_{ii}&=E_{+}\,,\,\,\,\,\,\,\,\,\,\Delta^{\textrm{eff}}_{ii}=0\,,\nonumber\\
h^{\textrm{eff}}_{ij}&=-\Delta\frac{e^{-r_{ij}/\xi_0}}{k_F r_{ij}}\sin(k_Fr_{ij})\langle\uparrow,i|\uparrow,j\rangle\,,\nonumber\\
\Delta^\textrm{eff}_{ij}&=\Delta\frac{e^{-r_{ij}/\xi_0}}{k_F r_{ij}}\cos(k_F r_{ij})\langle\uparrow,i|\downarrow,j\rangle\,.
\label{heff}
\end{align}
Here, $r_{ij}=|\textbf r_i-\textbf r_j|$ is the distance between impurities, $k_F$ is the Fermi wave vector, and $\xi_0=v_F/\Delta$ (with $v_F$ being the Fermi velocity) is the coherence length. \sh{The matrix elements, $h^\textrm{eff}_{ij}$ and $\Delta^\textrm{eff}_{ij}$, are obtained by calculating the corresponding overlaps $\langle \Psi_{+i} | H_C  | \Psi_{-j}\rangle$ and $\langle \Psi_{+i} | H_C | \Psi_{+j}\rangle$, see Refs.~\cite{pientkaPRB14}~and~\cite{hoffmanPRB16} for more details.}

\begin{figure*}[t!]
	\centering
	\includegraphics[width =0.65\textwidth]{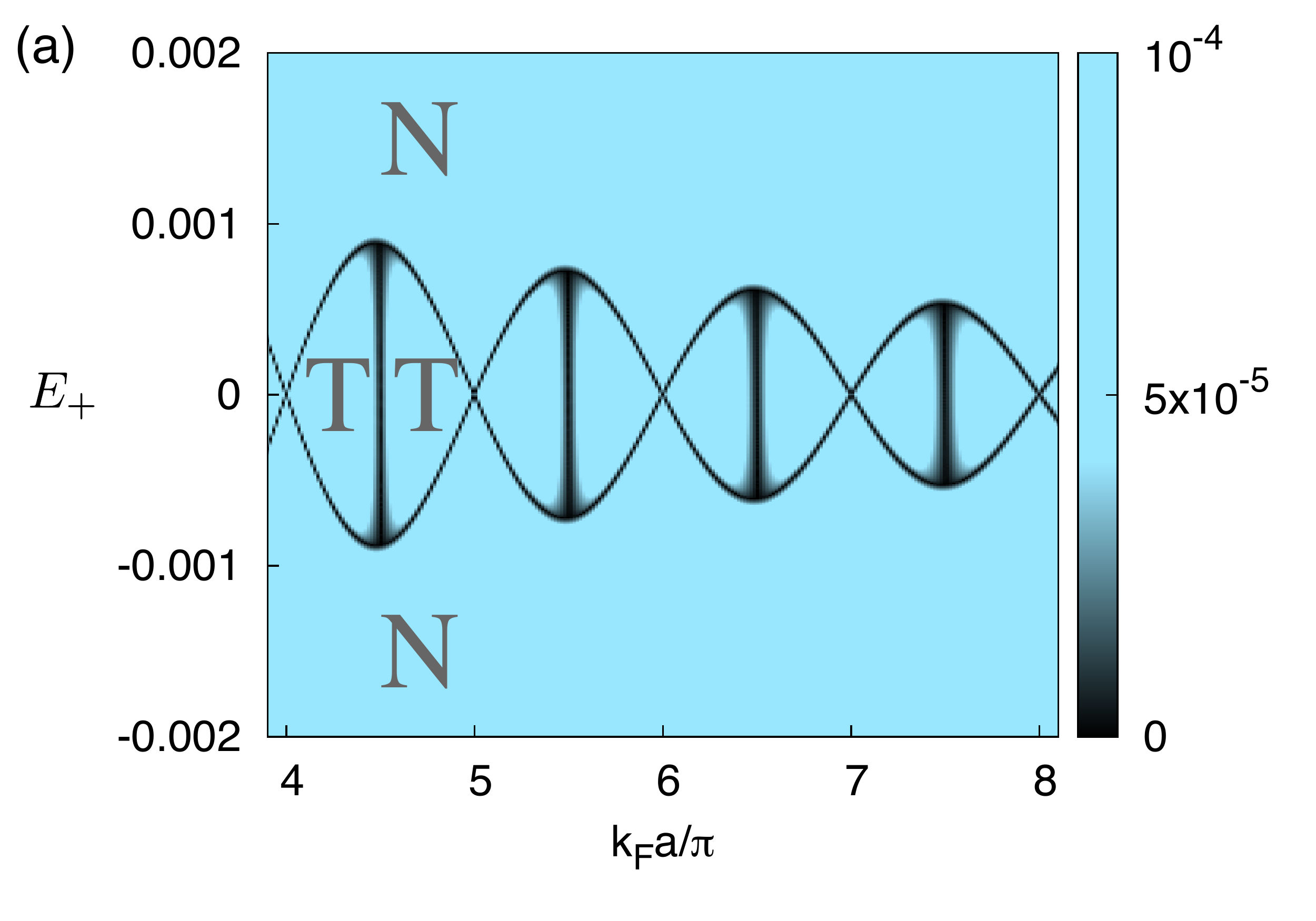}
	\caption{\jk{Magnitude of the smallest gap in the bulk spectrum as a function of the Fermi wavevector $k_F$ and energy level $E_+$ of the uncoupled Shiba state when the coherence length $\xi_0$ is small compared to the  spacing $a$. The topological (T) and nontopological (N) phases are separated by the contour $E_+=|\Delta_{ii+1}|$.} \rp{Reprinted figure with permission from Pientka {\it et al.} Topological superconducting phase in helical Shiba chains. {\it Phys. Rev. B} {\bf 88,}  155420 (2013).  Copyright 2013 by the American Physical Society.}}
	\label{phase1}
\end{figure*}

We immediately notice the similarities with Eq.~(\ref{Kitaev}) where the spin polarized Shiba states take the roll of the spinless fermions in the Kitaev chain. The chemical potential,  hopping between sites, and unconventional pairing amplitude are replaced by $E_+$, $h^{\textrm{eff}}_{ij}$, and $\Delta^\textrm{eff}_{ij}$, respectively. The critical difference is that, in general, the effective Hamiltonian includes longer-ranged hopping and superconducting pairing which can reach beyond nearest-neighbor sites (but it is bounded by the coherence length $\xi_0$).

\jk{In the limit where the coherence length of the superconductor $\xi_0$ is much shorter than the spacing between impurities, $r_{ii+1}=a$,  Eq.~(\ref{Heff}) effectively reduces to a nearest-neighbor Hamiltonian. As such, simply upon inspection and comparison with the Kitaev model, the condition for the Shiba chain to be in the topologically nontrival phase is that $E_+<h_{ii+1}$ and $\Delta_{ii+1}>0$ such that the system is gapped. As the topological and nontopological phases are separated by the closing of the gap, one can diagonalize Eq.~(\ref{Heff}) and find for which parameters the gap in the bulk spectrum closes thus determining the borders of the topological phase. Figure ~\ref{phase1} shows the smallest bulk gap as a function of the variables $E_+$ and $k_F$ and thus a \textit{de facto} phase diagram. As expected, the gap closes when $E_+=|h_{ii+1}|$, and this contour determines the boundary between the topological and nontopological phases. The gap also closes when $k_F a$ is an integer multiple of $\pi$ because the superconducting pairing is zero [see Eq.~(\ref{heff})].}

%In the limit where coherence length of the superconductor  $\xi_0$ is much shorter than the spacing between impurities,  Eq.~(\ref{Heff}) effectively reduces to a nearest-neighbor Hamiltonian. As such, simply upon inspection and comparison with the Kitaev model, the condition for the Shiba chain to be in the topologically nontrival phase is that $E_+<h_{ii+1}$ and $\Delta_{ii+1}>0$ such that the system is gapped. One can generate a phase diagram as a function of the variables $E_+$ and $k_F$ (see Fig.~\ref{phase1}).

\begin{figure*}[b!]
	\centering
	\includegraphics[width =0.65\textwidth]{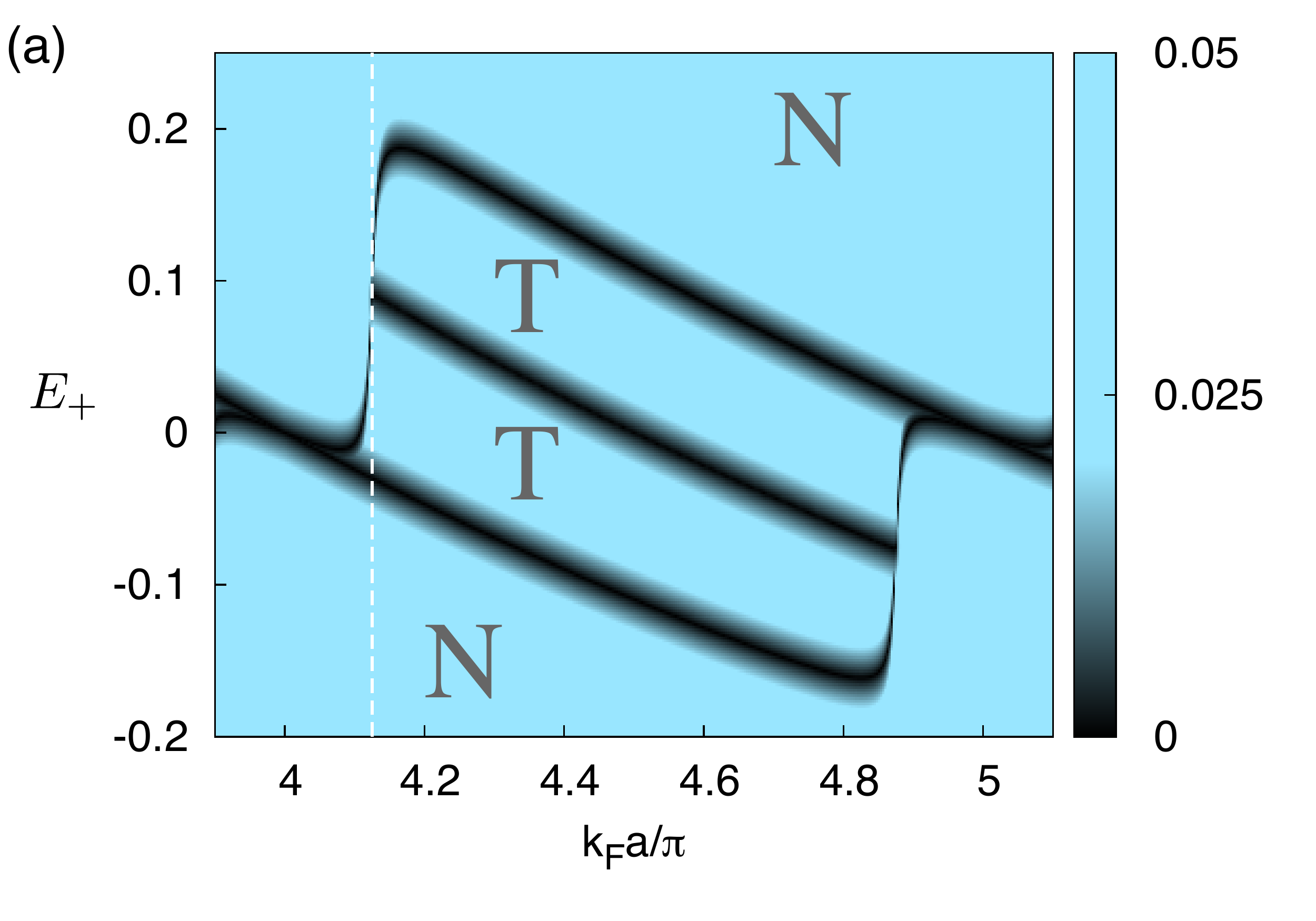}
	\caption{ \jk{Magnitude of the smallest gap in the bulk spectrum as a function of the Fermi wavevector $k_F$ and energy level $E_+$ of the uncoupled Shiba state when the coherence length $\xi_0$ is large compared to the spacing $a$ and $1/k_F$. The topological (T) and nontopological (N) phases are separated by the diagonal contours $E_+=-[k_F a (\textrm{mod}~2\pi)]/ k_F a$ and the condition $4\pi+k_h a<k_Fa < 5\pi-k_h a$. 
%	Energy level $E_+$ of the uncoupled Shiba state when the coherence length $\xi_0$ is small compared to the spacing $a$.
	} \rp{Reprinted figure with permission from Pientka {\it et al.} Topological superconducting phase in helical Shiba chains. {\it Phys. Rev. B} {\bf 88,}  155420 (2013).  Copyright 2013 by the American Physical Society.}}
	\label{phase2}
\end{figure*}

\jk{When $\xi_0$ is much longer than the spacing between impurities, both the normal hopping and superconducting pairing are power-low suppressed by $1/k_F r_{ij}$. Nonetheless, one can Fourier transform the effective Hamiltonian. Although the bulk spectrum is significantly more complicated \cite{pientkaPRB13}, the size of the gap in the bulk spectrum can be found and as well as when the gap closes. One finds a typical phase diagram (see Fig.~\ref{phase2}) in which the the vertical boundaries of the phases are defined by $n\pi+k_h a<k_Fa < (n+1)\pi-k_h a$ for an integer $n$ and the diagonal boundaries are defined by the contour $E_+=-[k_F a (\textrm{mod}~2\pi)]/ k_F a$. The topological phases have been corroborated by finding the MZM wave functions numerically \cite{pientkaPRB13} and analytically \cite{hoffmanPRB16}  as well as calculating the winding number of the bulk spectrum \cite{pientkaPRB14}.}

\subsection{Inducing superconductivity in a helical chain}
\label{wire}
When the electron hopping between atoms, controlled by the overlap of their orbitals, is much larger than the magnetic exchange interaction, the chain of atoms on a superconductor can be modeled as a one-dimensional wire with a proximity-induced superconductivity. Furthermore, the Ruderman-Kittel-Kasuya-Yosida (RKKY) \cite{rudermanPR54,kasuyaPTP56,yosidaPR57}  interaction is known to stabilize the magnetic order as helix with pitch twice the Fermi wave vector of the effective wire \cite{brauneckerPRL09,brauneckerPRB09,brauneckerPRB10}.
As such, we model the chain as a quantum wire  \cite{klinovajaPRL13,vazifehPRL13,brauneckerPRL13}:
\begin{align}
H=&H^{\textrm{kin}}+H^Z+H^{sc}\,.
\label{Hf}
\end{align}
The contributions to the Hamiltonian corresponds to the kinetic energy, magnetic exchange interaction, and proximity induced superconductivity are given by
\begin{align}
&H^\textrm{kin}=\sum_{\sigma} \int dr\  \Psi_\sigma^\dagger (r)\left[\frac{(-i \hbar \partial_r)^2}{2m}-\mu\right]\Psi_{\sigma} (r)\,, \nonumber\\
&H^Z=\Delta_m \sum_{\sigma, \sigma'} \int dr \ \Psi_\sigma^\dagger (r)\left[\cos(2k_F r)\sigma_x +\sin(2k_Fr)\sigma_y\right]_{\sigma \sigma'}\Psi_{\sigma'} (r)\,,\nonumber\\
&H^{sc}=\frac{\Delta}{2}\sum_{\sigma, \sigma'}\int dr\ [ \Psi_\sigma(r)(i\sigma_y)_{\sigma\sigma'} \Psi_{\sigma'}(r)+\textrm{H.c.}]\,,
\end{align}
respectively, where $m$ is the effective electron mass, $\mu$ is the chemical potential of the wire, $\Delta$ is the proximity-induced superconducting order parameter, and $\Delta_m$ is the energy splitting energy between two spin directions as a result of the magnetic exchange interaction. The operator $\Psi_\sigma$ ($\Psi^\dagger_\sigma$)  destroys (creates) an electron with spin $\sigma=\uparrow,\downarrow$, and $\sigma_i$ are the Pauli matrices acting in spin space. 

The Hamiltonian can be linearized around $\pm k_F$ by expressing the electron operators in terms of slowly varying left and right movers \cite{brauneckerPRB10,klinovajaPRB12,klinovajaPRL13},  $\Psi_\sigma(r)=R_\sigma(r)e^{ik_Fr}+L_\sigma(r)e^{-ik_Fr}$ to obtain two groups of modes, $\psi_r=(R_\uparrow,L_\downarrow,R_\uparrow^\dagger,L_\downarrow^\dagger)$ and $\psi_l=(L_\uparrow,R_\downarrow,L_\uparrow^\dagger,R_\downarrow^\dagger)$, of opposite helicity written in Nambu space. The components of the Hamiltonian density, $\mathcal H$, are written in terms of  $\psi=(\psi_r,\psi_l)$ as
\begin{equation}
H=\frac{1}{2}\int dr\ \psi^\dagger \mathcal H \psi =  \frac{1}{2}\int dr\ \psi^\dagger (\mathcal H^{\textrm{kin}}+ \mathcal H^{\textrm{sc}}+ \mathcal H^Z)\psi\,,
\end{equation} where we introduce
\begin{align}
\mathcal H^{\textrm{kin}}&=(-i\hbar v_F\partial_r)\sigma_z\otimes\mathbb1\otimes\rho_z\,,\nonumber\\
\mathcal H^{Z}&=\Delta_m\sigma_x\otimes\eta_z\otimes(1+\rho_z)/2\,,\nonumber\\
\mathcal H^{sc}&=\Delta\sigma_y\otimes\eta_y\otimes\mathbb1\,,
\label{lin}
\end{align}
after dropping all fast oscillating modes where $\psi=(\psi_r,\psi_l)$ and $v_F$ is the Fermi velocity. Here, $\eta_i$ and $\rho_i$ are the Pauli matrices acting in Nambu space and mode space (\textit{e.g.}, $\rho_x$ exchanges $\psi_r$ and $\psi_l$), respectively.

In order to find the zero-energy bound states, we specify the geometry of the wire to be semi-infinite: terminating it at $r=0$ and extending it for $r>0$. We can readily find zero energy eignestates of Eq.~(\ref{lin}) that are normalizable in this region. Transforming from the basis of left and right mover operators back to the fermion operator basis, $(\Psi_\uparrow,\Psi_\downarrow,\Psi_\uparrow^\dagger,\Psi_\downarrow^\dagger)$, the corresponding zero energy eigenstates are rewritten as
\begin{align}
\Phi^\pm_1&=\left( \begin{array}{c} -i\textrm{sgn}(\Delta-\Delta_m)e^{i k_F r}\\e^{-i k_F r}\\i\textrm{sgn}(\Delta-\Delta_m)e^{-i k_F r}\\e^{i k_F r}\end{array} \right) e^{-\kappa^- r}\,\,\,\,\,,
\Phi^\pm_2=\left( \begin{array}{c} e^{i k_F r}\\-ie^{-i k_F r}\\e^{-i k_F r}\\ie^{i k_F r}\end{array} \right) e^{-\kappa^+ r }\,,
\nonumber\\
\Phi^\pm_3&=\left( \begin{array}{c} ie^{-i k_Fr}\\e^{i k_Fr}\\-ie^{i k_Fr}\\e^{-i k_Fr}\end{array} \right)e^{-\kappa r}\,,\,\,\,\,
\Phi^\pm_4=\left( \begin{array}{c} e^{-i k_Fr}\\ie^{i k_Fr}\\e^{i k_Fr}\\-ie^{-i k_Fr}\end{array} \right)e^{-\kappa r}\,.
\end{align}
Here, $\kappa^\pm=|\Delta-\Delta_m|/v_F$ and $\kappa=\Delta/v_F$. According to our geometry, the wave function must vanish at $r=0$ in order to satisfy the boundary conditions. This is only possible if $\Delta_m>\Delta$ wherein one can compose a MZM as a linear superposition
\begin{equation}
\Phi_M=\left( \begin{array}{c} ie^{i k_F r}\\e^{-i k_F r}\\-ie^{-i k_F r}\\e^{i k_F r}\end{array} \right) e^{-\kappa^- r}-\left( \begin{array}{c} ie^{-i k_Fr}\\e^{i k_Fr}\\-ie^{i k_Fr}\\e^{-i k_Fr}\end{array} \right)e^{-\kappa r}\,.
\label{maj_sol}
\end{equation}
This solution carries two important messages \cite{klinovajaPRL13}: (1) there are two exponential decay lengths characterizing the MZM and (2) the probability density of the MZM, generally oscillates with the wavevector $2 k_F$, i.e. the period of oscillations is given by $\pi/k_F$.

\section{Experimental signatures of Majorana zero modes (MZM)}

In this section we detail the experimental efforts to observe MZMs at the ends of magnetic chains. All the experiments have in common the utilization of scanning tunneling microscopy (STM) measurements which provide a spectral resolution together with a direct real-space visualization of the MZMs. This is in contrast to transport measurements made in, for instance, quantum wire systems which have access only to the spectral properties \cite{Mourik2012,Deng2012,Churchill2013,Das2012,Finck2013,Albrecht2016,Guel2018}. The differences within these magnetic chain experiments come from the how the chains are built, either using self-assembly or individual atomic manipulation, and additional measurement tools, such as spin-polarized scanning tunneling microscopy (SP-STM) and atomic force microscopy (AFM).

%The most recent proposal in attempting to detect MZMs in solid-state systems is likely the atomic chains made of magnetic atoms on $s$--wave superconductors~\cite{Nadj-Perge2013}. In comparison with conductance measurements of semi-conducting nanowires hosting MZMs, such atomic chains investigated by STM/STS have the great asset to provide a spectral resolution together with a direct real-space visualization of the MZMs.  Due to the chain geometry and in contrast to the topological insulator/superconductor hetero-structures, both MZMs at chain ends are in principle accessible by STM/STS.

\label{MajoSTM}

\subsection{Experimental signature of Shiba states}

\begin{figure*}[t!]
	\centering
	\includegraphics[width =0.65\textwidth]{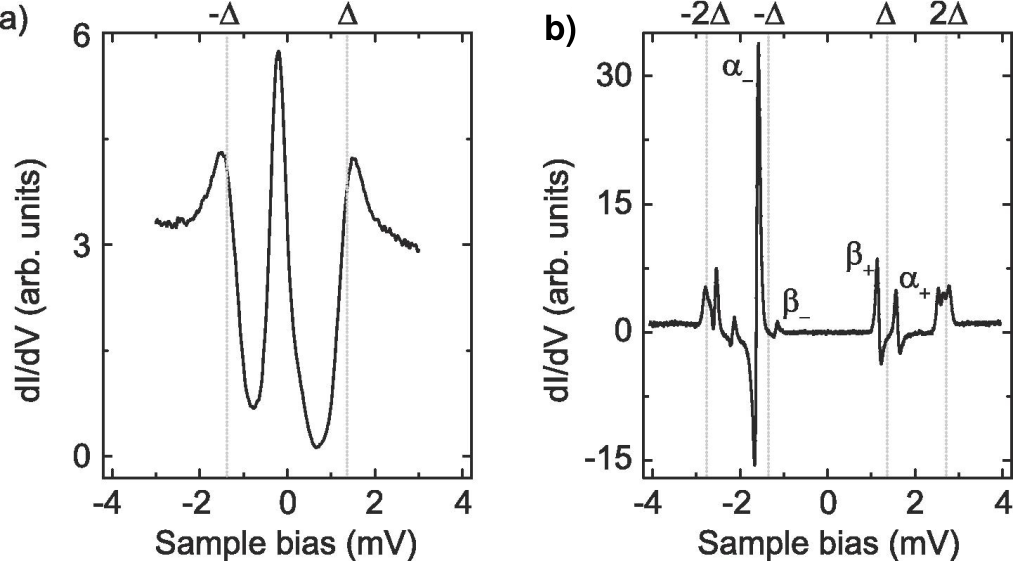}
	\caption{{\bf a,} Differential conductance spectra $dI/dV$ of Yu-Shiba-Rusinov (YSR) states of Mn atom on Pb(111)  acquired with a metal tip at 1.2 K. \rp{Reprinted from Heinrich {\it et al.} Single magnetic adsorbates on s-wave superconductors. {\it Prog. Surf. Sci.} {\bf 93} 1--19 (2018). Copyright 2018 with permission from Elsevier.} {\bf b,}  
	Differential conductance spectra $dI/dV$ of Yu-Shiba-Rusinov (YSR) states of Mn atom on Pb(111)  acquired with a Pb-covered superconducting tip at 1.2 K. Different numbers of Shiba resonances are resolved depending on the tip material and the temperature. \rp{Reprinted figure with permission from Ruby {\it et al}. Tunneling processes into localized subgap states in superconductors. {\it Phys.
Rev. Lett.}  {\bf 115} 087001 (2015). Copyright 2015 by the American Physical Society.}.}
	\label{ShibaSTM}
\end{figure*}

Before delving into MZMs, we briefly mention the experimental evidence for \rp{Shiba states obtained by scanning tunneling microscopy (STM)~\cite{Yazdani1997,Ruby2015a,Heinrich2018}.} Figure~\ref{ShibaSTM}a is \rp{a differential conductance spectra obtained with the lock-in technique,} $dI/dV$,  of a Mn atom on Pb(111) with a metallic tip at 1.2 K, whereas Fig. \ref{ShibaSTM}b is acquired using a superconducting tip on the same atoms under the same conditions. When the STM tip is in normal state, the energy resolution of the Shiba states is limited by the thermal broadening of the metallic electrons from the tip ($\sim 300 \mu V$ at 1.2 K). The Shiba state induced by a magnetic impurity on a superconductor then appears as a single peak as shown in Fig.~\ref{ShibaSTM}a~\cite{Yazdani1997}. Using a superconducting tip, this limitation is lifted  leading to an increase of the sub-gap states resolution~\cite{Ji2008,Ji2010,Ternes2006,Franke2011,Heinrich2013,Hatter2015,pengPRL15,Chevallier2016}. %\jk{Y. Peng, F. Pientka, Y. Vinkler-Aviv, L. I. Glazman, and F. v. Oppen, Phys. Rev. Lett. 115, 266804 (2015).}
Since the spectral function is obtained by convoluting tip and sample density of states (DOS), the resonance peak positions in the superconducting gap are shifted by $\pm \Delta_{\textrm{tip}}$ with $\Delta_\textrm{tip}$  the superconducting gap of the tip. The subgap resonances from the Shiba states with energy $\alpha$ symmetric to $E_F$ (thus $\alpha^+$ and $\alpha^-$ in Fig.~\ref{ShibaSTM}b) then appear at $E = \pm (\Delta_\textrm{tip} + \alpha)$. The energy resolution is of order $\sim$ 50 $\mu eV$ at 1.2 K.

%%%%%%%%%%%%%%%%%%%%%%%%%%%%%%%%%%%%%%%%%%
%%%%%%%%%%%%%%%%%%%%%%%%%%%%%%%%%%%%%%%%%%%%%%%%%%%%%%%%%%%%%%%%%%%%%%%%%%%%%%%%%%%%%%% 

With STM spectroscopy, not only the spectral resolution of Shiba states can be experimentally acquired but also their spatial localizations. As evident from Eq.~(\ref{Heff}) and Eq.~(\ref{heff}), the Shiba wavefunction decays as power in the Fermi wavelength $2\pi/k_F$ and exponential in the superconducting coherence length away from the impurity. Remarkably, experiments have reported such a decay of the Shiba wavefunction realized by $\textrm{NbSe}_2$ and of Mn adatoms on Pb surfaces \cite{Menard2015,Ruby2016}.

\subsection{MZM signatures in atomic chains by scanning tunneling spectroscopy (STS)}

\begin{figure*}[ht!]
	\centering
	\includegraphics[width =0.80\textwidth]{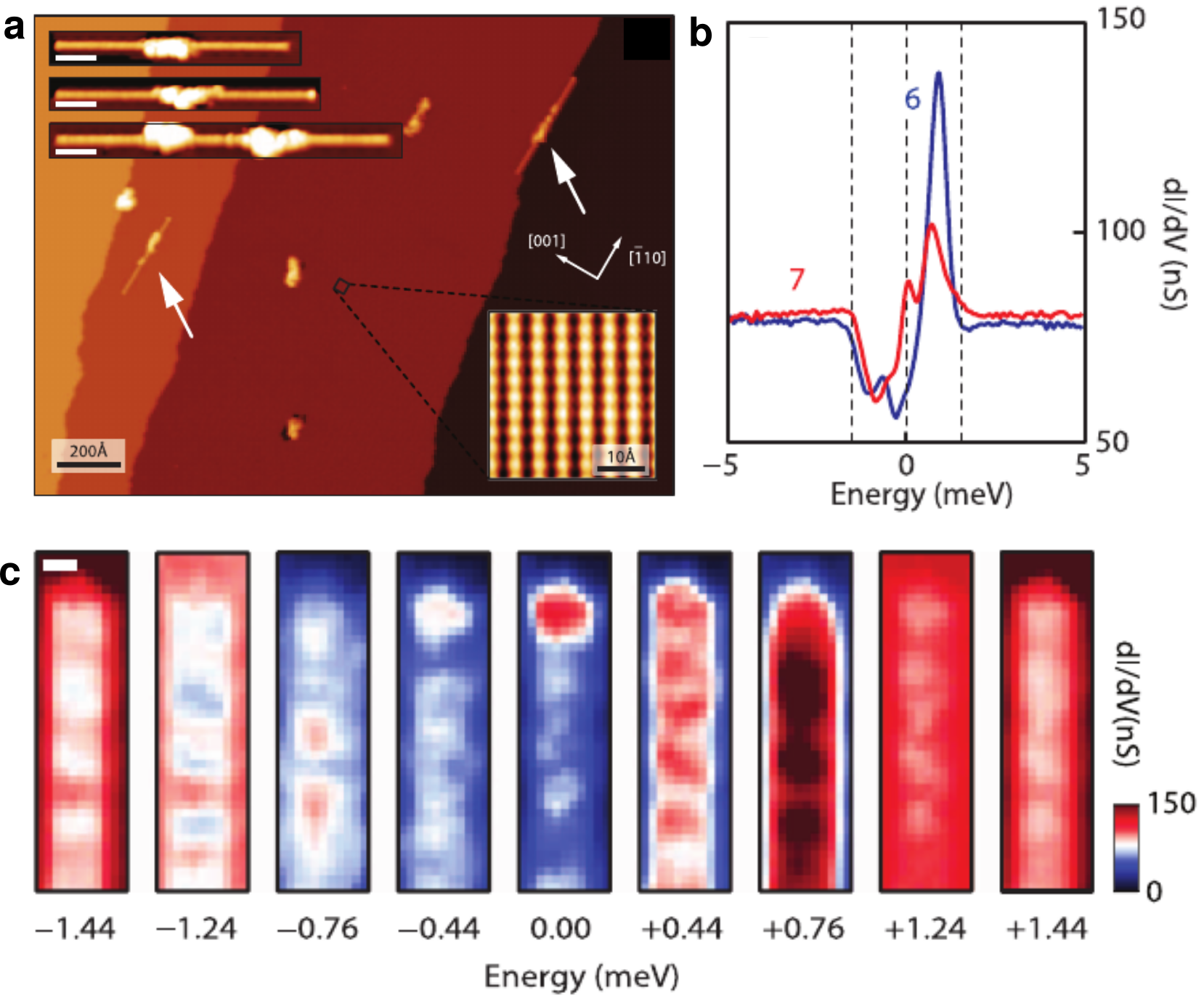}
	\caption{Zero-bias conductance peaks in magnetic chains on superconducting lead. {\bf a,} STM topography of the Fe atomic chains on Pb(110) indicated by white arrows and atomically clean terraces of Pb (see inset, scale bars are 5 nm). {\bf b,}  Single-point differential conductance spectra, $dI/dV$, acquired in the middle of the chain (blue) and at its end (red). The ZBCP is a signature of  MZMs. {\bf c,} $dI/dV$ maps of the chain end taken at various voltages. At zero-energy, the ZBCP is localized at the chain end as predicted by theory (scale bar is 1 nm). \rp{From S. Nadj-Perge {\it et al.} {\it Science} {\bf 2014,} 346 602--607. Reprinted with permission from AAAS.}} 
	\label{FigYaz}
\end{figure*}

The first experimental implementation of magnetic atomic chains featuring zero-bias peaks at the ends was achieved by Nadj-Perge {\it et al.} in 2014~\cite{Nadj2014}. The system was obtained by self-assembly of Fe atoms on Pb(100) under ultrahigh vacuum conditions. 
An STM image (see Fig.~\ref{FigYaz}a) shows such a Fe chain of $\sim$ 20 nm in length on the Pb surface. As predicted by theory (see Sec.~\ref{Kitaevsection}), MZMs are expected to appear at the chain ends when entering the phase of topological superconductivity, which does not require any fine-tuning in this particular case.  
Such MZMs provide a conducting channel at zero energy in the otherwise insulating subgap region. Thus, to check for the presence of MZMs, local $dI/dV$ spectra acquired at the chain end (red) and at the middle (blue) with a metallic tip were compared as shown in Fig.~\ref{FigYaz}b. The zero-bias conductance peak (ZBCP), being taken as evidence for an MZM, is detected at the chain end and vanishes at its middle (blue curve) in agreement with theory. 
Figure~\ref{FigYaz}c shows $dI/dV$ maps acquired at different energies within the SC gap (between $\pm$ 1.44 eV).  At $E$ = 0 eV, the ZBCP (red contrast) is localized at the very end of the chain, together with a non-negligible local density of state (LDOS) throughout the chain. At other energies, contributions from other mid-gap states are also detected since the local density of state (LDOS) is relatively strong all along the chain. 
   
%%%%%%%%%%%%%%%%%%%%%%%%%%%%%%%%%%%%%%%%%%%%%%%%%%%%%%%%%%%%%%%%%%%%%%%%%%%%%%%%%%%%%
%%%%%%%%%%%%%%%%%%%%%%%%%%%%%%%%%%%%%%%%%%%%%%%%%%%%%%%%%%%%%%%%%%%%%%%%%%%%%%%%%%%%%%%    
   
\begin{figure*}[t!]
	\centering
	\includegraphics[width =0.80\textwidth]{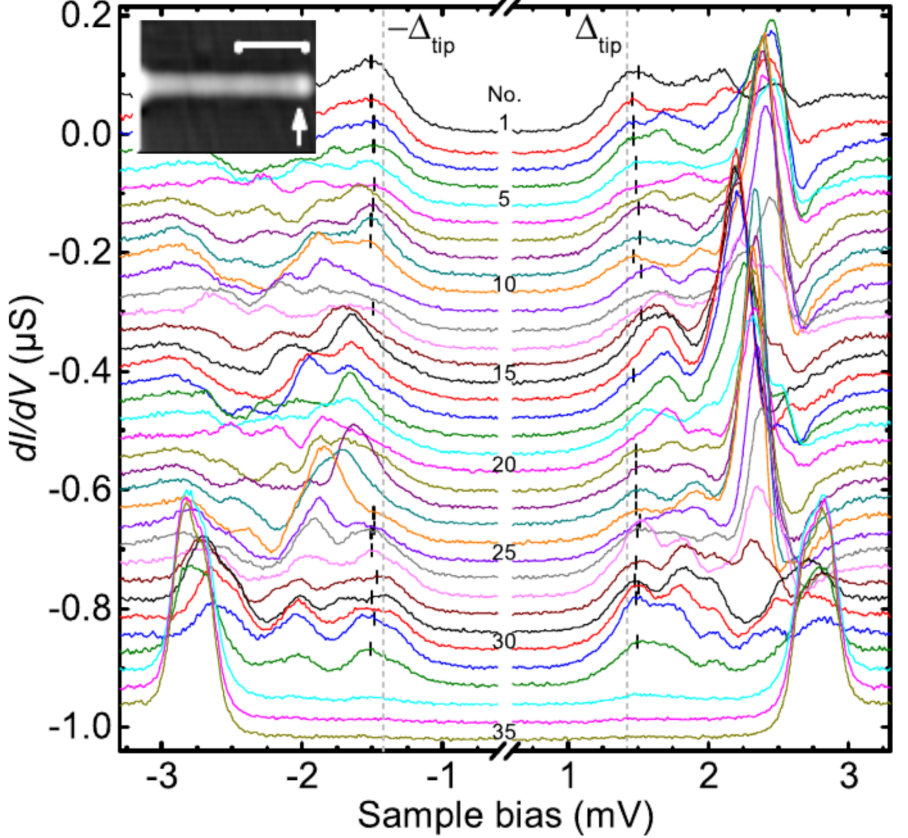}
	\caption{Finite-bias conductance peaks in magnetic chains on superconducting lead. Spatially-resolved differential conductance, $dI/dV$, along the Fe chain on Pb(111) (see inset) acquired with a superconducting tip. At the chain end (curve 29), a zero-energy conductance peak emerges at +$\Delta_{tip}$ = 1.42 meV, indicating an MZM.
\rp{Reprinted figure from M. Ruby {\it et al.} {\it Phys. Rev. Lett.} {\bf 115} 197204 (2014). Copyright 2014 by the
American Physical Society.}} 
	\label{FigFranke}
\end{figure*}

As discussed in Sec.~\ref{shiba}, the DOS contributions captured by these $dI/dV$ maps likely come from subgap Shiba states due to the magnetic Fe atoms on Pb(110). In Ref.~\cite{Ruby2015}, Ruby {\it et al.} further characterized this aspect by reproducing the differential conductance measurements with superconducting STM tips to readily increase the spectral resolution~\cite{,Ji2008,Ji2010,Ternes2006,Franke2011,Heinrich2013,Heinrich2018}. Figure~\ref{FigFranke} shows single-point $dI/dV$ spectra acquired along a Fe chain on Pb(111) (displayed in the inset). The graph is composed of 35 curves with both extrema potentially hosting MZMs that correspond to curves No 1 and No 29. The $dI/dV$ spectra of the Pb(110) surface is in curve No 35. Since STS measurements with SC tips provides a convolution of the DOS of tip and sample, the sample resonances are shifted by the tip SC gap such as $E(eV) = \pm (\Delta_{tip} + \Delta_{sample})$. Zero-energy states in Fig.~\ref{FigFranke} are thus at $E_0 = \pm \Delta_{tip}$ = $\pm$ 1.42 meV.

For curves No 1 and 29, the ZBCPs indicating the presence of MZMs are visible either as a peak (+ 1.42 meV) or a shoulder (-1.42 meV). The absence of symmetry of the ZBCPs with respect to $E_F$ is explained by the authors as due to the  overlap with the nearby low-energy resonances which lift the particle-hole symmetry in the tunneling current. Comparing with other sub-gap resonances observed throughout the chain, the ZBCP does not change in energy neither in intensity. Note that further STS spectroscopic data acquired with superconducting tips at 20 mK were also provided by the group of Yazdani~\cite{Feldman2016} further revealing the presence of Shiba resonance peaks inside the superconducting gap.

%%%%%%%%%%%%%%%%%%%%%%%%%%%%%%%%%%%%%%%%%%%%%%%%%%%%%%%%%%%%%%%%%%%%%%%%%%%%%%%%%%%%%
%%%%%%%%%%%%%%%%%%%%%%%%%%%%%%%%%%%%%%%%%%%%%%%%%%%%%%%%%%%%%%%%%%%%%%%%%%%%%%%%%%%%%%%

An interesting output of the work of Ruby {\it et al.}~\cite{Ruby2015} is the accurate analysis of these sub-gap states with such high-spectral resolution as well as the observation of the resulting $d$-bands at much higher energies crossing the Fermi level near the chain end~\cite{Ruby2015}. These $d$--bands provide support for an alternative explanation of topological superconductivity in the Fe/Pb(110) system. In this scenario, such magnetic chains on $s$--wave superconductors require the formation of a spin-polarized $d$--band. Since the electron bands cross the Fermi level with an odd number and with the strong spin-orbit interaction of the Pb surface, a $p$--wave pairing in the magnetic chain is expected and thus a topological phase. Furthermore, and as shown in previous Sections, the disappearance of these ZBCPs at the chain ends was demonstrated by suppressing the superconducting state when increasing the sample temperature or increasing the magnetic field in accordance with the expected behavior of MZMs~\cite{Nadj2014,Ruby2015,Pawlak2016}.

\subsection{Spin texture of Fe and Co magnetic chains on superconducting Pb}

%%%%%%%%%%%%%%%%%%%%%%%%%%%%%%%%%%%%%%%%%%%%%%%%%%%%%%%%%%%%%%%%%%%%%%%%%%%%%%%%%%%%%
\begin{figure*}[t!]
	\centering
	\includegraphics[width =\textwidth]{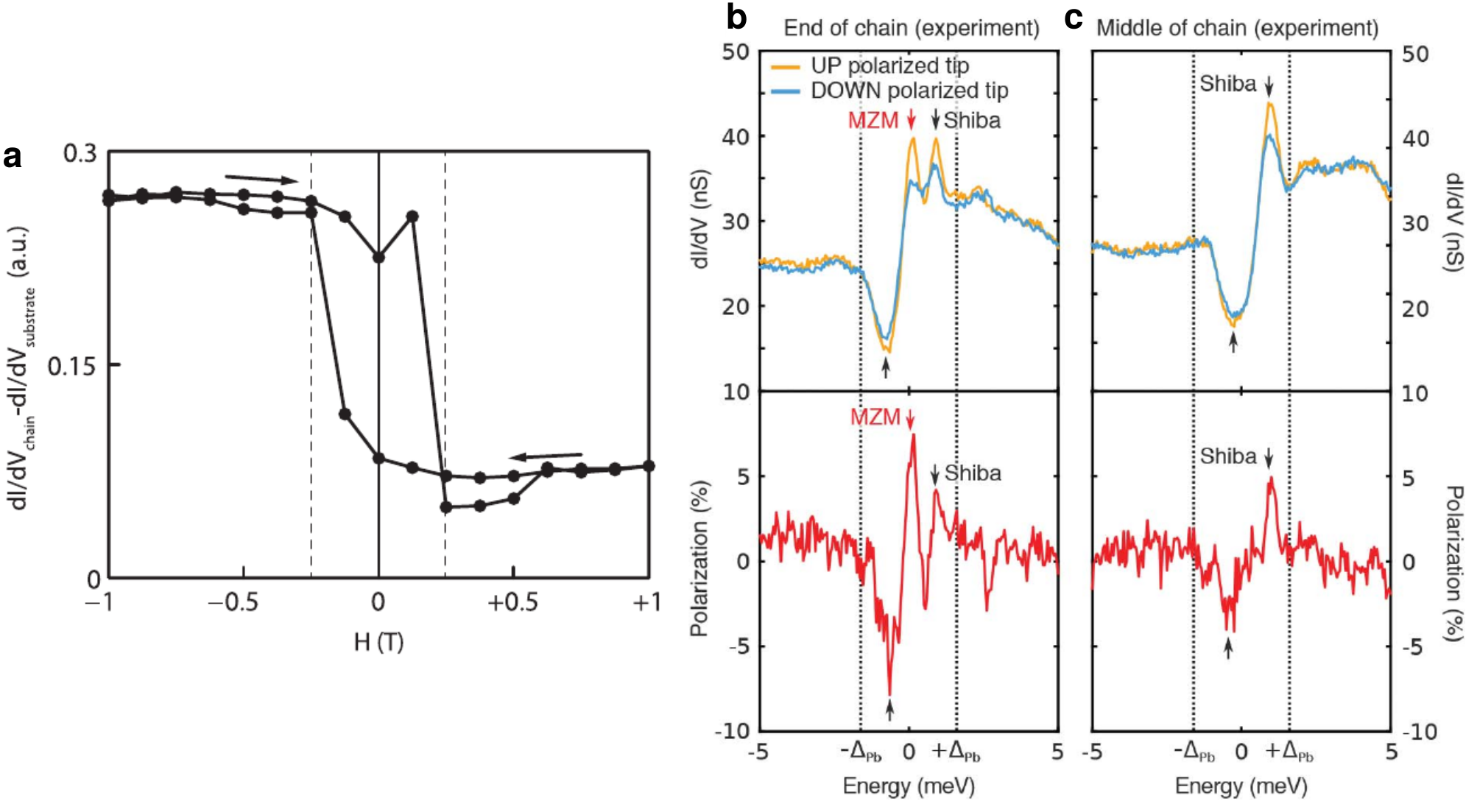}
	\caption{Spin texture of the Fe atomic chains on superconducting lead. {\bf a,} Difference of spin-polarized conductance between the Fe chain and the substrate measured with a Cr/Fe tip. \rp{From S. Nadj-Perge {\it et al.} {\it Science} {\bf  346} 602--607 (2014). Adapted with permission from AAAS.} {\bf b,} Spin-polarized differential conductance, $dI/dV$, at the end and the middle of the chain. Yellow and blue curves refer to spin up and down polarized tips. The MZM is marked with a red arrow. The bottom plots are the corresponding spin polarization obtained from the experimental spectra. \rp{From S. Jeon {\it et al.} {\it Science} {\bf 358} 772--776 (2017). Reprinted with permission from AAAS. }} 
	\label{Magnet}
\end{figure*}
%%%%%%%%%%%%%%%%%%%%%%%%%%%%%%%%%%%%%%%%%%%%%%%%%%%%%%%%%%%%%%%%%%%%%%%%%%%%%%%%%%%%%%%

The magnetic nature of the Fe chains on Pb(110) has been first characterized by Nadj-perge {\it et al.} using spin-polarized STM with Cr tips at 1.2 K~\cite{Nadj2014}. By comparing the spin-polarized $dI/dV$ measurements on top of the chains and on the Pb substrate while ramping the $B$ field (Fig.~\ref{Magnet}a), they observed a hysteresis loop over the chain, as a characteristic signature of a ferromagnetic ordering. The field switching was experimentally determined at $\sim$ 0.25 T. Further spin-polarized measurements from the same group~\cite{Jeon2017} conducted at lower temperature ($\approx$ 20 mK) focused on determining the spin-polarization of the sub-gap states and of the ZBCP (Fig.~\ref{Magnet}b-c). The spectroscopy data-sets acquired at the chain end and in the middle are shown in Figures~\ref{Magnet}b and c, respectively. Both spectra show the spin-polarization of a \rp{Shiba} band throughout the chain that is in agreement the observations of Ref.~\cite{Ruby2015} for Fe chains.  Concerning the MZM (red arrow in Fig.~\ref{Magnet}b), its spin--polarization is explained as given by the spin-polarization of the \rp{Shiba} band crossing $E_F$ at the chain ends.  

Parallel to this work, the group of Franke also tackled the characterization of self-assembled Co chain on Pb(110) by STM/STS with superconducting tips and spin-polarized STM at 1.2~K~\cite{Ruby2017}. Similar to the Fe case study, the Co adatoms self-assemble in a one-dimensional fashion on Pb(110) where the emergence of MZM is expected at their ends. Surprisingly and in stark contrast to the Fe chain, Ruby {\it et al.} showed the systematic absence of MZMs in these chains. %The replacement of Fe atoms by Co ones modifies the electronic configuration within the chain. It hovever still leads to sub-gap "spin-polarized" YSR resonances as scrutinized by spin-polarized STM. 
 Similar to Ref.~\cite{Jeon2017} for the Fe case, spin-polarized STM demonstrated the appearance of hybridized \rp{Shiba} states from the Co atoms of the chain resulting in spin-polarized $d$-bands. Although the spin-polarization of the chain in proximity to the $s$-wave superconductor (having a strong spin-orbit interaction) provides the fundamental background for topological superconductivity, the absence of MZMs in that case is explained by the difference of oxidation states of the Co atoms compared to the Fe case. Ruby {\it et al.} suggested that the band structure of the Co chain system exhibits an even number of Fermi points within half the Brillouin zone (in contrast to the odd number of the Fe case study~\cite{Ruby2015}) which precludes the transition to a topological phase and thus MZMs. 

At present, and in view of the experimental data provided in Refs.~\cite{Nadj2014,Jeon2017,Ruby2017},  there is a general 
consensus to conclude to the ferromagnetic nature of these self-assembled atomic chains on superconducting Pb(110). Nevertheless, we note that a  helical ordering of the chains with a small opening angle (i.e. with nearly ferromagnetic ordering) is still compatible with these data.
Within the Kitaev model, such ferromagnetic chains are predicted by theory to host MZMs at the chain ends without fine tuning, as discussed in Section~\ref{Kitaevsection} (if they have an odd number of Fermi points). 
We emphasize here that the original theoretical proposals considered a helical spin polarization from Ruderman--Kittel--Kasuya--Yosida (RKKY) interactions  to allow the MZM emergence~\cite{klinovajaPRL13,vazifehPRL13,brauneckerPRL13}.  [For MZM proposals generated by a linear disordered array of nanomagnets with randomly oriented magnetizations, see Ref.~\cite{Choy2011}. The periodic arrays of nanomagnets were considred in Refs. \cite{klinovajaPRL12,kjaergaardPRB12,PhysRevLett.117.077002}.
%\jk{J. Klinovaja, P. Stano, and D. Loss, Phys. Rev. Lett. 109, 236801 (2012).; M. Kjaergaard, K. Wolms, and K. Flensberg, Phys. Rev. B 85, 020503(R) (2012). }]
%Comments: ~\cite{Choy2011,Nadj-Perge2013,Kjaergaard2012,Klinovaja2013}. Note that Choy2011 does neither consider RKKY nor helical order, they %instead assume random orientation of magnetic moments of nanmoagnets. Similarly, Kjaergaard2012  does not consider a spin chain and not RKKY but  consider a rotating B field. This was considered before by Bernd Braunecker, George I. Japaridze, Jelena Klinovaja, and Daniel Loss. Phys. Rev. B 82, 045127 (2010) and it was shown there that that SOI and rotating (helical)  B field are equivalent.
We also note that such spin helices have been indeed observed in other 1D-atomic system by STM measurements~\cite{Menzel2012,Menzel2014,Steinbrecher2018,Kim2018}.

\subsection{Atomic structure of the Fe chain}
%%%%%%%%%%%%%%%%%%%%%%%%%%%%%%%%%%%%%%%%%%%%%%%%%%%%%%%%%%%%%%%%%%%%%%%%%%%%%%%%%%%%%
\begin{figure*}[t!]
	\centering
	\includegraphics[width =\textwidth]{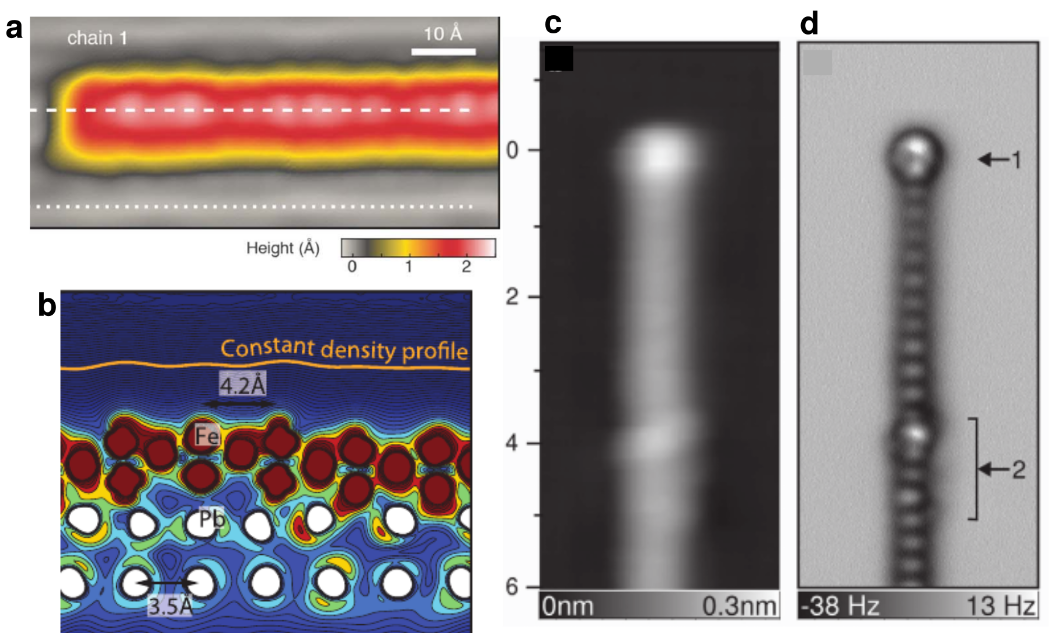}
	\caption{ Structure of the Fe chains on Pb(110). {\bf a,} Typical STM image of the Fe chain on Pb(110). The chain has an apparent height of $\sim$ 0.2 \AA~ and exhibits two periodic modulations of $\sim$ 15--18 \AA~and $\sim$ 3.5 \AA. From S. Jeon {\it et al.} {\it Science} {\bf 2017,} 358 772--776. Adapted with permission from AAAS. {\bf b,} Vertical cross-section of the relaxed structure by DFT showing a zig-zag type structure. The Fe atoms (red) are stacked vertically and embedded into the Pb surface. \jk{From S. Nadj-Perge {\it et al.} {\it Science} {\bf 346,} 602--607 (2014). Adapted with permission from AAAS.} {\bf c,} STM image and {\bf d,} the corresponding constant-height AFM image of a Fe chain on Pb(110). Each protrusion corresponds to a Fe atom, clealry visible in the AFM image  ({\bf d}). The Fe chain in {\bf d} is seen to be mono-atomic, linear, and does not show any vertical atomic corrugation as depicted in {\bf a}. The Fe-Fe distance in image {\bf d} is 0.37 nm in close agreement with the Pb surface lattice constant. In image {\bf d}, $1$ indicates the position of the ZBCP and $2$ refers to a slight misalignment of the Fe atoms along the chain. \jk{ Figures {\bf c} and {\bf d} adapted from R. Pawlak {\it et al.} {\it npj Quantum Info} {\bf 2,} 16035 (2016)}.} 
	\label{Struc}
\end{figure*}
%%%%%%%%%%%%%%%%%%%%%%%%%%%%%%%%%%%%%%%%%%%%%%%%%%%%%%%%%%%%%%%%%%%%%%%%%%%%%%%%%%%%%%%
From a fundamental point view, the chain structure has also a pivotal role in the interpretation of the experimental data. Until now, we have not discussed this aspect in detail which, however, mediates the electronic and magnetic coupling along the chains. In the aforementioned references, the atomic Fe chains, grown using very similar preparation procedures, appear linear as shown in the STM image Fig.~\ref{Struc}a with an apparent height of $\sim$ 2 \AA~ and centered between the atomic rows of the Pb(110). Two periodic modulations are always superimposed in images of $\sim$ 15--20 \AA~and 3.5 \AA, respectively~\cite{Nadj2014,Ruby2015,Pawlak2016}. At first glance, the exact position of the Fe atoms in the structure is difficult to interpret only taking into account the STM images. In Ref.~\cite{Nadj2014}, Nadj-Perge {\it et al.} proposed a structural model of the Fe chain system on Pb(110) based on their experimental observations and density functional theory (DFT). The DFT relaxed structure is presented in Figure~\ref{Struc}b. The calculations conclude to a strong Fe-Pb bonding leading to a chain structure which is a partially embedded zigzag chain of atoms between Pb(110) atomic rows. From this, the contour of the constant-electron density have been computed (orange line Fig.~\ref{Struc}b) and found consistent with the STM images reported in the literature~\cite{Nadj2014,Ruby2015,Pawlak2016}. 

%%%%%%%%%%%%%%%%%%%%%%%%%%%%%%%%%%%%%%%%%%%%%%%%%%%%%%%%%%%%%%%%%%%%%%%%%%%%%%%%%%%%%
\begin{figure*}[ht!]
	\centering
	\includegraphics[width =0.8\textwidth]{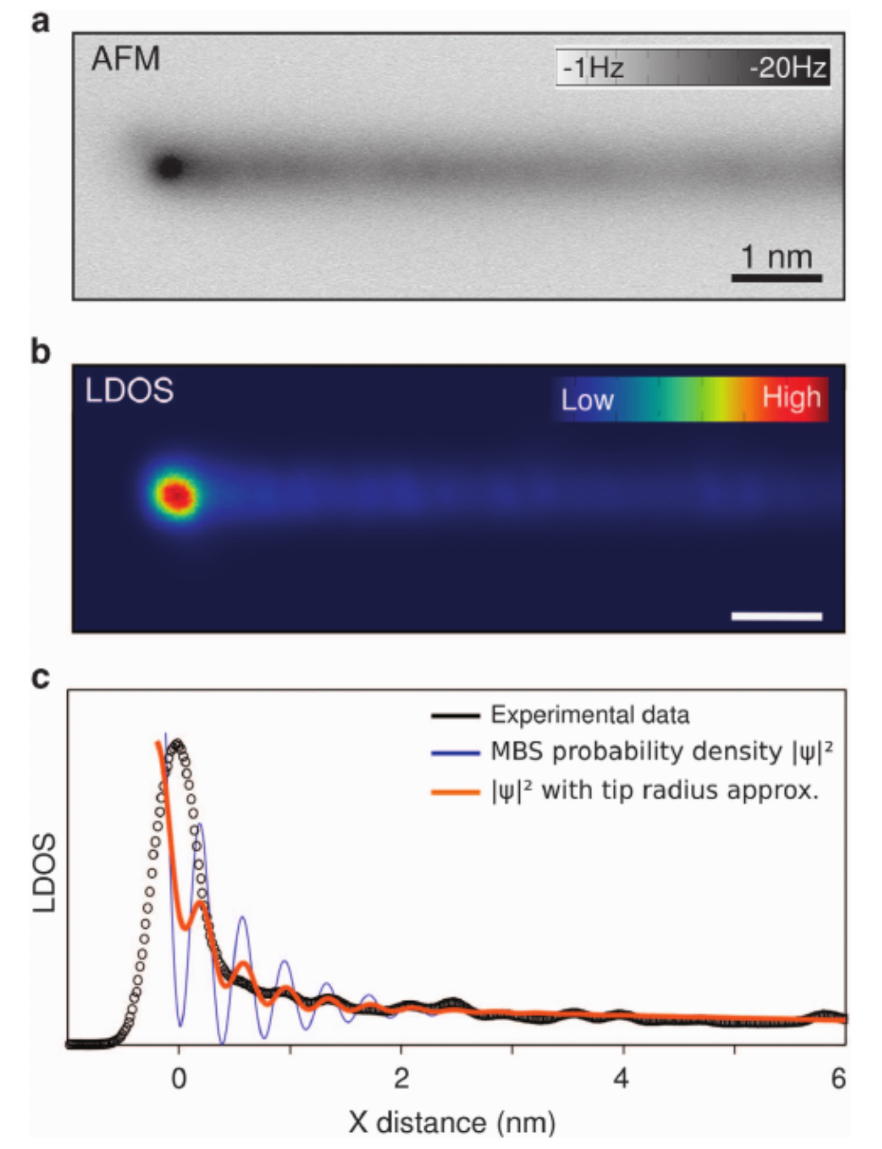}
\caption{\rp{Majorana localization lengths. {\bf a,} Constant-height AFM image and {\bf b,} Corresponding zero-bias $dI/dV$ maps of the chain end hosting a MZM. {\bf c,} $dI/dV(X)$ profile (black dots) taken along the chain revealing the localisation lengths of the zero-bias conductance peak ($X$ = 0). The blue curve is the probability density $|\psi|^2$ of a Majorana bound state with two localization lengths $\xi_1 \approx$ 22 nm and $\xi_2 \approx$ 0.72 nm. The orange curve approximates $|\psi|^2$ by considering the effect of the tip radius. $\xi_1$ and $\xi_2$ corresponds to 2 and 59 atomic sites, respectively. Adapted with permissions from Pawlak {\it et al.} Probing atomic structure and Majorana wavefunctions in mono-atomic Fe chains on superconducting Pb surface. {\it npj Quantum Information.} {\bf 2} 16035 (2016).} }
	\label{PAW}
\end{figure*}
%%%%%%%%%%%%%%%%%%%%%%%%%%%%%%%%%%%%%%%%%%%%%%%%%%%%%%%%%%%%%%%%%%%%%%%%%%%%%%%%%%%%%%%

Using the recent advances in high-resolution imaging with AFM~\cite{Gross2009,Pawlak2011,Pawlak2012,Kawai2016}, Pawlak {\it et al.} characterized the structure of Fe chains on Pb(110) at 4.2~K~\cite{Pawlak2016}. Figures~\ref{Struc}c and d show an STM image of such a chain and the corresponding constant-height AFM image, respectively. From the AFM data showing each Fe atom as a protrusion,  the Fe-Fe inter-atomic distance along the chain was found to be $\sim$ 0.37 nm, close to the Pb surface lattice (0.35 nm) and much shorter than the Fe-Fe period denoted in Fig.~\ref{Struc}b. This AFM observation suggests a high commensurability of the chain period with the substrate and a lattice mismatch of $\sim$ 0.6 $\%$. Note also that the AFM imaging is highly sensitive to atomic corrugation since it has proven, for instance, the detection of height variations in single molecules within $\sim$ 0.05 \AA~precision~\cite{Kawai2016,Pawlak2019}. As a result, the atomic buckling of $\sim$ 0.65 \AA~along the triple zig-zag chain should be easily reflected in the AFM images, which instead, show flat and linear atomic chains (Fig.~\ref{Struc}d). Particular slightly misaligned Fe atoms can be observed along the chain (position $2$), whereas at then end of the chain (position $1$) there is a halo in the AFM image at exactly the location where the ZBCP was measured by STS. 

An interesting outcome of extracting the exact chain structure is the comparison of the localization length of the ZBCP measured by the STS with the theoretical predictions~\cite{klinovajaPRL13,Chevallier2016}. By comparing the decay of the ZBCP from the $dI/dV$ profiles as function of chain sites (Fig.~\ref{PAW}), one can determine that the decay of the ZBCP peak is characterized by two localization lengths \cite{Pawlak2016} [see Eq. (\ref{maj_sol})]: 
\begin{eqnarray}
 \xi_1 = \frac{\hbar v_F}{\Delta}\,, \, \,\,\,\,  \xi_2 = \frac{\hbar v_F}{\Delta_m - \Delta},
 \end{eqnarray}
where $\Delta$ is the superconducting proximity gap in the chain, $\Delta_m$ is the effective Zeeman field produced by the helically ordered magnetic moments of Fe in the chain, and $k_F$ = 8.3 nm$^{-1}$ is the Fermi wavevector. \sh{As a result, the authors extracted a short localization length, $\xi_1 = 7.5$ \AA ~($\sim$ 2 atomic sites) and a long one, $\xi_2$, extending up to 220 \AA~($\sim$ 60 atomic sites) ((Fig.~\ref{PAW}c), which is still relatively short compared to MZM localization lengths in nanowires.} The helical gap associated with $\xi_2$ is found to be $\Delta_m \approx$ 33 meV. This suggests that the relatively short localization length of the MZM in such hybrid Fe/Pb systems arises from weak exchange intreaction between itinerant electrons in the Fe chain and the localized d-shell electrons of the Fe atoms playing the role of the magnetic impurity spin~\cite{klinovajaPRL13}. 

\subsection{Atom-by-atom construction of Fe chains on superconductors}
%%%%%%%%%%%%%%%%%%%%%%%%%%%%%%%%%%%%%%%%%%%%%%%%%%%%%%%%%%%%%%%%%%%%%%%%%%%%%%%%%%%%%
\begin{figure*}[ht!]
	\centering
	\includegraphics[width =0.9\textwidth]{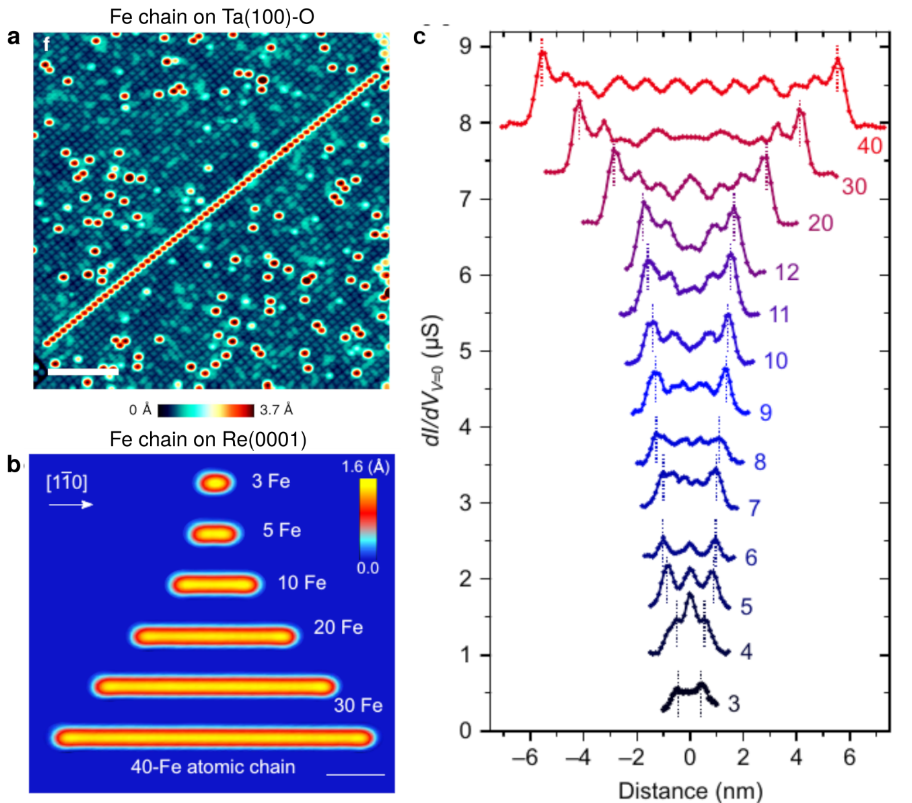}
\caption{Atom-by-atom constructed Fe chains on superconductors. {\bf a,} STM image of a long chain consisting of 63 Fe atoms on the oxidized Ta(001) in the superconducting state (T = 1.2 K). The scale bar is 10 nm.  \rp{Due to the weak and inhomogenous exchange coupling between Fe atoms within the chains and with the substrate, this system does not develop a Shiba band throughout the chain and thereby lacks the formation of MZMs at the chain ends.} \rp{Adapted with permissions from Kamlapure {\it et al.} Engineering the spin couplings in atomically crafted spin chains on an elemental superconductor. {\it  Nat. Comm.} {\bf 9} 3253 (2018).} {\bf b,} STM images of the artificially constructed Fe chains of various lengths on superconducting Re(0001). The chains exhibit hybridized Shiba bands and a helical spin polarization leading to topological superconductivity. {\bf c,} Experimental zero-energy $dI/dV$ profiles along the chains for 3 to 12, 20, 30, and 40 Fe atoms (from bottom to top). A ZBCP indicating the presence of MZMs is measured at the chain ends. \rp{Adapted with permissions from Kim {\it et al.} Toward tailoring majorana bound states in artificially constructed magnetic atom chains on elemental superconductors. {\it Sci. Adv.} {\bf 4} eaar5251 (2018).}
}
	\label{Atombyatom}
\end{figure*}
%%%%%%%%%%%%%%%%%%%%%%%%%%%%%%%%%%%%%%%%%%%%%%%%%%%%%%%%%%%%%%%%%%%%%%%%%%%%%%%%%%%%%%%
To avoid structural defects in the self-assembled chains and gain control over the chain structure, the atom-by-atom construction by tip-induced manipulations is a solid alternative~\cite{Crommie1993,Pawlak2017}. The first attempt to realize this using an STM on superconducting surface was reported by Kamlapure {\it et al.} in 2018~\cite{Kamlapure2018}. They used as superconductor, the $(3\times3)$ oxygen reconstructed Ta(100) surface (named Ta(100)-O, critical temperature $T_c \approx 4 K$, SC gap $\approx$ 0.63 meV) on top of which they deposited single Fe atoms. Through successive manipulation processes consisting of vertical atomic displacements, a chain of 63 atoms was formed over the surface (Fig.~\ref{Atombyatom}a). By comparing the STS spectral signature at 1.2 K obtained by \rp{using Ta-coated superconducting tips of single Fe atoms and atoms within constructed chains~\cite{Kamlapure2018}, they demonstrated the presence of Shiba resonances within the superconducting gap $\Delta_{SC}$ = 1.26 meV [$\Delta_{SC} = \pm (\Delta_{tip} + \Delta_{sample})$ with $E_0 = \pm \Delta_{tip}$ = $\pm$ 0.63 meV]} as well as their slight shifts in energy while increasing the number of atoms of the chain. \rp{ From the analysis of \rp{Shiba} peak positions, the authors concluded to unequivalent exchange coupling of the Fe atoms and the substrate as well as a weak short-range interaction bewteen neigboring Fe atoms along the chain.} \rp{ However, in contrast to the Fe/Pb(110) system~\cite{Nadj2014,Ruby2015,Feldman2016,Jeon2017,Ruby2017}, such a chain does not develop a hybridized Shiba band along the system which excludes the formation of MZMs at the chain ends.}

This tip-induced manipulation method was then used by the same group to construct Fe chains of various lengths on another superconductor, i.e. Re(0001) \cite{Kim2018}. Due to its low transition temperature of $\sim$ 1.6~K, all measurements were  carried out at much lower temeprature (300 mK). The STM image of Fig.~\ref{Atombyatom}b shows the artificially built-up Fe chains along the [1$\overline{1}$0] direction. \rp{Here, similar to Fe and Co chains on Pb(110)~\cite{Nadj2014,Ruby2015,Jeon2017,Ruby2017}, $dI/dV$ spectra showed clear signatures of sub-gap Shiba states in the superconducting gap ($\Delta_{Re}$ = 0.28 meV) and a hybridized Shiba band throughout the chain.} The corresponding spin-polarized STM imaging obtained with magnetic tips further showed clear spatial modulations with periods of $\sim$ 3.6 \AA~  and $\sim$ 11 \AA~(corresponding to 4/3$a_{Re}$ and 4$a_{Re}$ with $a_{Re}$ the Re lattice parameter), respectively. Using Fe-coated Pt/Ir tips sensitive to either the out-of-plane or in-plane spin components, they observed a sligth shift of the modulation position with respect to the chain. Thus, the authors concluded from this shift depending on the tip-polarization to the presence of a spin spiral originating from Dzyaloshinsky--Moriya interaction~\cite{Menzel2012,Menzel2014}. 

Strinkingly, similar spin-polarized STM measurements of the Fe chains on Pb(110) also exhibit two spatial modulations with similar periods ($\sim$ 3.5~\AA ~and $\sim$ 15--18~\AA)~\cite{Jeon2017} as for the Fe chain on Re(0001). In this case, since no changes of the modulation position with different spin--polarization of the tips were observed, a ferromagnetic character of the chain was concluded and the modulations explained by the triple zig-zag structure~(Fig.~\ref{Struc}b)~\cite{Nadj2014}. 

Figure~\ref{Atombyatom}c shows zero-energy $dI/dV(X)$ profiles being proportional to the LDOS acquired along chains varying from 3 to 40 atoms (Fig.~\ref{Atombyatom}b) in order to estimate the localization length of the ZBCP at each end (if any) and when they start to hybridize and split away from zero energy. Most profiles are symmetric with respect to the center of the chain with a spatial modulation along the chains. The minimum chain size to obtain a well--localized ZBCP at the chain end is found to be  $\approx$ 12 atoms and more. Below this value, the LDOS at the chain end and in the middle are very similar which could be explained by strongly hybridized MZMs. 

\section{Open questions}

The experiments presented in the previous section conclusively observe zero-bias peaks in the STM measurements at the ends of the chain of magnetic atoms which is evidence in support of the presence of MZMs. The mechanism that drives the system into a topological phase remains, however, under dispute. In this section, we explicate these differences in the interpretation of the data and propose tests to resolve them.

\subsection{Magnetic order of the chain}
\label{order}

In the theoretical analysis of the data, the authors of Ref. \cite{Pawlak2016}  assumed a helical magnetic ordering in the chain. Upon making a local gauge transformation, one can `untwist' the helix to obtain a ferromagnetic order and a spin-dependent gauge field, \textit{i.e.} a Rashba spin-orbit interaction~\cite{brauneckerPRB10}. That is, observation of the MZMs as well as the shape of their wavefunctions at the chain ends does not shed light on the magnetic ordering in the magnetic chain: the outcome of measurements will be the same for both ferromagnetic and helical orderings if the chemical potential is inside the topological gap, which is automatically the case for the self-tuning RKKY mechanism.
  However, the order itself as well as the mechanism for the order is important to (1) fully understand the physics of the system and (2) to engineer new atomic structures.

There are three dominant competing mechanisms for the magnetic order. The first is the RKKY interaction \cite{rudermanPR54,kasuyaPTP56,yosidaPR57} which typically describes the ordering of magnetic impurities due to their exchange interaction with conduction electrons in metals or with quasiparticles in  superconductors. In superconductors, in the absence of any spin-orbit interaction, the coupling between two magnetic impurities oscillates between ferromagnetic and antiferromagnetic. However, for many impurities, the competition between magnetic order and superconductivity forces the chain of impurities into a helix. When the impurities are embedded in a three dimensional superconductor (Sec.~\ref{shiba}), the pitch of the helix is proportional to $(k_F^{2}/\xi)^{1/3}$, where $k_F$ and $\xi$ are the Fermi wavevector and superconducting coherence length, respectively  \cite{andersonPR59,abrikosovBK88}. When superconductivity is proximity-induced in a one-dimensional chain (Sec.~\ref{wire}), the pitch of the helix is $2k_F$, where $k_F$ is the Fermi wavevector in the proximitized chain~\cite{klinovajaPRL13,brauneckerPRL13}. \sh{The magnitude of the spin susceptibility is proportional to $1/(k_F\xi)^{2/3}$ in the former case and proportional to $1/k_F a$ (for the spacing $a$ between adatoms) in the latter case. Therefore, for a sufficiently dilute (dense) arrangement of the magnetic impurities, $\xi\ll a$ ($\xi\gg a$), one would expect the 3D (1D) pitch to dominate. Thus, in a one-dimensional Shiba chain in the dense limit, the RKKY interaction is effectively one-dimensional. In additon, if the coupling between the adatoms of the chain is stronger than the coupling between them and the superconducting substrate, the 1D RKKY limit holds. However, to complicate matters even further, if there is additionally a spin-orbit interaction, the spin symmetry of the standard RKKY interaction gets broken and a helix with pitch proportional to the magnitude of the spin-orbit interaction can be favored~\cite{hoffmanPRB16,heimesNJoP15}. }
%
%	RKKY interaction in carbon nanotubes and graphene nanoribbons
%Jelena Klinovaja and Daniel Loss.
%Phys. Rev. B 87, 045422 (2013)
%
%RKKY Interaction On Surfaces of Topological Insulators With Superconducting Proximity Effect
%Alexander A. Zyuzin and Daniel Loss.
%Phys. Rev. B 90, 125443 (2014)
%
%Transition from fractional to Majorana fermions in Rashba nanowires
%J. Klinovaja, P. Stano, and D. Loss.
%Phys. Rev. Lett. 109, 236801 (2012).

In addition to the delocalized quasiparticles in superconductors, the presence of the localized Shiba states can likewise affect the preferred magnetic order of the chain. As quasiparticles states largely outnumber the Shiba states, the former often determine the ordering. However, resonant Shiba states, \textit{i.e.} $E_+\approx0$, can overcome the effect of the quasiparticles \cite{yaoPRL14} and switch the magnetic ordering of two impurities from ferromagnetic to antiferromagnetic and vice versa. Furthermore, it is known that when $E_+<0$, the superconductor undergoes a quantum phase transition in which, local to the impurity, the magnitude of the superconductivity dramatically decreases and changes sign \cite{schlottmannPRB76,salkolaPRB97,balatskyRMP06,flattePRL97,flattePRB97,bjornsonPRB15}. 
 This inhomogeneity in the superconducting order parameter can itself change the favored magnetic orientation \cite{hoffmanPRB15}. 

Finally, using a simple symmetry argument, one can see that by placing the atomic chain on top of the superconducting substrate can induce an easy-axis anisotropy which favors a ferromagnetic or antiferromagnetic orientation. The magnitude, direction, and sign of this anisotropy are specific to the details of how impurities interact with the surface of the superconductor.

In order to distinguish the order of the magnetic chain, we propose applying a magnetic field in the plane of the superconducting substrate, forcing the the magnetic moments of impurities to align along that direction. If the MZMs remain, this would imply that the helical order is not important to the system. 

\subsection{Where does the MZM live?}

As discussed in the theory section, there are two distinct scenarios that could support MZMs: (1) when the magnetic exchange interaction is large, the Shiba states form a band which mimics a long-distance analog of the Kitaev chain and (2) when orbitals of the magnetic atoms hybridize and proximity superconductivity is induced in an effectively one-dimensional channel. 

The evidence for the former was originally proposed in Ref.~\cite{Nadj2014}. The STM measurements showed a corrugation along the magnetic chain. After running density functional theory calculations (DFT) modeling Fe atoms on the surface of superconducting Pb, they found that strong bonding between the atoms resulted in chain structure partially embedded in the Pb, resulting in a zigzag Fe structure, dipping in and out of the Pb along the chain. This further implied that the exchange interaction of these ferromagnetic chains is of $\sim2.4$~eV. Note that such ferromagnetic character has been also concluded from the spin-polarized measurements in Refs.~\cite{Ruby2015,Ruby2017}. Although the exchange interaction is rather large, of the order of the Fermi energy of Pb, it implies that the Shiba states are near  zero energy, $E_+\approx0$, and thus, according to the phase diagrams (see Figs.~\ref{phase1} and \ref{phase2}), capable of supporting MZMs at the ends of the Shiba chain. 

In a subsequent experiment \cite{Pawlak2016}, the AFM was used to complement the STM measurements. Using the substantial resolution of the AFM, the authors showed that their magnetic chain sample is linear and does not have a zigzag structure which obviated the necessity for a large magnetic exchange. Additionally, by probing the signature for MZMs, the authors measured the oscillations and decay of the wavefunction which was matched with the theoretical wavefunction of an MZM formed within the atomic chain [see Eq.~(\ref{maj_sol})]. As a result, they extracted the Fermi wavevector, magnetic exchange interaction, and the proximity-induced superconducting gap in the Fe chain, $k_F=8.3$~nm$^{-1}$, $\Delta_m=33$~meV, and $\Delta=1.1$~meV, respectively. As $\Delta<\Delta_m$,
% and \sh{the Fermi energy of the Pb is much larger than the Fermi energy in the Fe chain}, 
 this interpretation is consistent with MZMs formed in the chain, and rather supports the scenario of a helical spin-polarization as recently observed experimentally~\cite{Kim2018}.  

To distinguish between these two scenarios, we propose adding an additional, nonmagnetic atom to the end of the chain which can act as two level quantum dot. Being nonmagnetic, it will not form a bound state within the superconductor. As a result, if the MZM is formed at the ends of the Shiba band, it will be largely decoupled from the end mode. However, if the MZM is formed directly in the chain, the MZM could still leak into the bulk superconductor \cite{vernekPRB14,leak1,leak2} and a finite weight of the MZM could be observable there. \rp{Interestingly, recent experiments have already shown that a substantial spectral weight of the MZMs leaks to the Pb atoms adjacent to the Fe chain ends~\cite{Feldman2016}. This results in the observation in $dI/dV$ maps of a ``double-eye" feature close to the MZM location as well as the detection of a ZBCP from buried Fe chains through Pb overlayers. Our proposal could be experimentally realized by, for instance, coupling copper adatoms to the ends of the prototypical Fe chains on Pb(110).} 

\sh{\subsection{Non-Abelian statistics and braiding of MZMs in magnetic chains}
Although the experimental evidence presents a compelling case for the existence of zero energy bound states at the ends of magnetic chains (see Sec.~\ref{MajoSTM}), it has not distinguished the results from other zero energy modes, such as Kondo resonances \cite{chengPRX14}, Andreev states \cite{liuPRB17,ptokPRB17,moorePRB18,vuikCM18,avilaCM18,aseevPRB18,fleckensteinPRB18,reegPRB18}, or Tamm-Shockley states \cite{tammPZS32,shockleyPR39}. In order to uniquely identify the end states as MZMs, one needs to consider the effect of exchanging three MZMs to determine if these states indeed obey non-Abelian statistics.}

In the following, we outline how to exchange two MZMs which can be immediately generalized to the exchange of three MZMs. The typical setup to braid two MZMs consists of two 1D topological superconductors oriented perpendicular to each other and such that the end of one coincides with the middle of the other, forming a so-called `T-junction' \cite{Alicea2011}. Crucially, one must have spatially resolved control of the topological phase over the entirety of the T-junction. Invoking this control, the T-junction is tuned so that the horizontal part of the T is in the topological phase and the vertical part is in the trivial phase such that MZMs reside only at the two ends of the horizontal chain. Starting from one end, we choose the right end for concreteness, and moving to the left, the sites are adiabatically tuned out of the topological phase thereby effectively moving the right MZM across the horizontal portion of the T-junction. Once the MZM reaches the junction of the two topological superconductors, the vertical part is adiabatically tuned into the topological phase starting from the junction and moving vertically down until the MZM is at the end of the vertical topological superconductor. Performing the same procedure, the left MZM is moved to the right side and the formerly right MZM, residing at the bottom of the T, is moved to the left side of horizontal topological superconductor. Thus, the MZMs have been exchanged. Upon attaching an additional T-junction supporting another pair of MZMs to the existing one, it is easy to generalize the braiding of two MZMs to three.

\sh{In nanowires, the topological phase can be manipulated using backgates to effectively control the local chemical potential. As such gates are experimentally difficult to implement in nanostructures such as those discussed in the previous section, magnetic chains do not at the moment have the luxury of voltage controlled topological phase. Furthermore, it is unclear if such a change in the local chemical potential could induce a topological phase transition.}

\sh{Alternatively, we propose applying a local magnetic field to control the topological phase. For a sufficiently large magnetic field, this has a twofold effect. First, it will locally (within the spatial reach of the magnetic field) polarize the magnetic impurities into a ferromagnetic orientation. Second, it will change the energy of the Shiba states. If helical order is indeed responsible for inducing the $p$-wave superconductivity, the local ferromagnetic order  will either demote the superconducting pairing to an $s$-wave type as in the nanowire setup (see Sec.~\ref{wire}) or will destroy the superconducting pairing outright as in the Shiba chain (see Sec.~\ref{shiba}). Thus, this is expected to change the topological phase. If the impurity ordering is ferromagnetic to begin with, the additional magnetic field can drive the Shiba states away from the chemical potential, analogous to the effect of a local chemical potential on electronic states. In the first case the magnetic field must be strong enough to overcome the anisotropy responsible for the helical order, while in the second case the magnetic field must be comparable to the Fermi energy; we estimate a local magnetic field of the order 1~T \cite{klinovajaPRL13} which may be induced by, for instance, a magnetic tip \cite{Jeon2017}. In either case, we believe such a local magnetic field could be an effective tool in moving, or even creating, MZMs. Alternatively, one can play with the direction of applied fields, however, this would require ordering magnetic adatoms in an atomic ring \cite{Li2016}.
}

\sh{\subsection{Topological Qubits from Magnetic Chains}
As discussed in Sec.~\ref{Kitaevsection}, two MZMs at the ends of a topological superconductor can be formally combined into a complex fermion whose state can either be occupied or unoccupied. Although this furnishes a two-level system, it is not possible to construct a superposition of states with different particle numbers and thus not sufficient to support the basis for a qubit. Rather, one minimally needs two topological superconductors. A convenient, two-level system is defined by restricting to an odd total number of fermions shared between the topological superconductors. The two states are defined by the occupancy of one of the two topological superconductors while the other remains unoccupied. Critically, in order to avoid decoherence, the two levels must be degenerate. This can be achieved by making sure that the two topological superconductors share the same chemical potential, which is the case if they are connected by a metallic bridge.
%An important and principal advantage to using magnetic chains to furnish MZM qubits is the atomic precision by which they can be constructed; the two states of the qubit are expected to be degenerate with a high degree of precision. This is in contrast to, for instance nanowires, which can vary in size, induced superconductivity, spin-orbit interaction, etc., all of which can split the ideally degenerate states. \jk{I am not sure I get your point here}
}

\sh{One may braid the MZMs, as described in the previous subsection, to perform some of the necessary gates for universal quantum computation \cite{Kitaev2003,freedmanCiMP02,bravyiAoP02,freedmanBAMS03}. Although operating the qubit in such a way is relatively error-free, it is difficult to perform experimentally and does not guarantee the necessary gates for a fully functional quantum computer. As such, a fashionable alternative is to use an ancillary, nontopological system to perform readout and operations while using the MZMs as a robust memory \cite{bravyiPRA06,leijnsePRL11,leijnsePRB12,hyartPRB13,hoffmanPRB16b,pluggePRB16,karzigPRB17}. In nanowires, this is enabled using a quantum dot \cite{leijnsePRL11,leijnsePRB12,hoffmanPRB16b,pluggePRB16,karzigPRB17}, typically defined within the nanowire \cite{dengSCI16}, whose coupling to the MZM can be controlled by using a local chemical potential to increase or decrease the barrier between the two \cite{denis12}. In addition, such a quantum dot can be also used as a detection tool, allowing to address the spin polarization of the bulk modes in the chain \cite{pawel1,spinmf1}.
}

\sh{In magnetic chains, the natural analogue to a quantum dot, i.e. a quasi-zero-dimensional level, could be provided by additional atoms on top of the superconductor. Such an atom could hybridize with the MZMs via a direct orbital overlap with the chain or, for a magnetic atom, by creating a Shiba state which can similarly overlap with the hybridized Shiba states created by the chain. Although this potentially provides a level, using such a level to perform quantum operations requires dynamic control of the coupling between the MZMs and the levels of the additional atom \cite{leijnsePRL11,leijnsePRB12,hoffmanPRB16b}; a theoretical or experimental understanding of how to control such a coupling in magnetic chains is unknown. However, we note that if one had local control of the levels of the Shiba states, as proposed in the previous section, quantum dot-like islands could be isolated within the same magnetic chain that supports the MZMs, analogous to quantum dots in nanowires. That is, consider a magnetic chain for which we had local control of the Shiba states energies by, for instance, a local magnetic field. Let one part of the chain be tuned to the topological phase and one part in the trivial phase. In the nontopological part but near one of the MZMs, one can create a small region with a single level. The barrier between the wave function of the single level and the MZM can be increased or reduced by changing the local magnetic field thereby dynamically changing the tunneling between the states. This proposal could provide a possible route to universal quantum computation supported by MZMs.}

\section{Conclusion and outlook}
The MZM is the solid--state counterpart of the Majorana particle in high-energy physics~\cite{majorana37}. Since the first experimental signatures in proximitized semiconducting nanowires~\cite{Mourik2012}, six years of intense researches have brought a plethora of experimental reports in diverse condensed-matter systems together with new theoretical interpretations and scenarios. The engineering of topological superconductors is the forerunner of the Majorana quest in experiments.  To date and beside the focus of this review on magnetic chains, signatures of MZMs have been also observed by conductance measurements at ultra-low temperatures in semiconducting nanowires~\cite{Mourik2012,Lutchyn2018}, at the interface of topological insulators/superconductors heterostructures~\cite{Xu2014,Xu2015,Sun2016,Sun2017,Wang2014} as well as in``natural" topological superconductors~\cite{Guan2016,Lv2017}. In agreement with theory, the most fundamental characteristics of the MZM have been observed in experiments, i.e. a robust zero-energy character accompanied by signatures of topological superconductivity~\cite{Mourik2012,Wang2018}, the localization of MZMs at the ends of atomic chains~\cite{Nadj2014,Ruby2015,Pawlak2016,Kamlapure2018}, a relatively short localization length of the MZM wavefunction~\cite{Albrecht2016,Pawlak2016,Kamlapure2018}, and a $2e^2/h$ quantized conductance by transport measurements~\cite{Zhang2018}. Altogether, these observations lend reasonable support to the existence of MZMs in solid-state systems. At the same time, an alternative explanation that the observed zero- bias peaks in transport measurements could also arise due to topologically trivial Andreev bound states originating within the normal regime has been put forth \cite{liuPRB17,ptokPRB17,moorePRB18,vuikCM18,avilaCM18,aseevPRB18,fleckensteinPRB18,reegPRB18}.

MZMs in solid--state systems are believed to be topologically protected against perturbations and exhibit non-Abelian exchange statistics which could serve as future $qubits$ for topological computations~\cite{Kitaev2003}. (However, we note that braiding of MZMs alone cannot provide all necessarily quantum gates and needs to be supplemented by additional non-topologically protected operations \cite{bravyiPRA06,hoffmanPRB16b,landauPRL16,pluggePRB16,karzigPRB17}.) 
 The demonstration of  braiding operations with MZMs will require beforehand the design of atomically--precise nano-structures with desired topologies and gates in these extreme conditions to enable MZM manipulations and the study of their interactions. These experimental advances would open new insights into the exotic physics of MZMs and their non-Abelian statistics, providing 'smoking-gun' evidence of MZMs. 

\section*{References}
\bibliography{Bibfile}

\end{document}